\documentclass[%reprint,
preprint,
 amsmath,amssymb,
 aps,
]{revtex4-2}

\usepackage{graphicx}% Include figure files
\usepackage{dcolumn}% Align table columns on decimal point
\usepackage{bbm}
\usepackage{bm}% bold math
\usepackage{booktabs}
\usepackage{amsmath}
\usepackage{nicematrix}

\definecolor{oxfordblue}{rgb}{0.0, 0.13, 0.28}

\begin{document}

%\preprint{PRX Life}

% \title{Theory of Aster Positioning and Dynamics}
%\title{Extensions and Analysis of the Stoichiometric Model for Centrosome Dynamics}
%\title{Centrosome Dynamics by Cortical Pulling Forces}
\title{A first-principles geometric model for dynamics of motor-driven centrosomal asters}

\author{Yuan-Nan Young$^{1,3}$}
\author{Vicente Gomez Herrera$^{2}$}
\author{Helena Z. Huan$^{2}$}
\author{Reza Farhadifar$^{3}$}
\author{Michael J. Shelley~\footnote{Corresponding author: mshelley@flatironinstitute.org}$^{ ,2,3}$}
\affiliation{$^{1}$ Department of Mathematical Sciences, New Jersey Institute of Technology, Newark, New Jersey 07102, USA}
\affiliation{$^{2}$ Courant Institute, New York University, NY, NY 10012, USA}
\affiliation{$^{3}$ Center for Computational Biology, Flatiron Institute, NY, NY 10010, USA
}

\begin{abstract}
The centrosomal aster is a mobile and adaptable cellular organelle that exerts and transmits forces necessary for tasks such as nuclear migration and spindle positioning. Recent experimental and theoretical studies of nematode and human cells demonstrate that pulling forces on asters by cortically anchored force generators are dominant during such processes. Here we present a comprehensive investigation of a first-principles model of aster dynamics, the S-model (S for stoichiometry), based solely on such forces. The model evolves the astral centrosome position, a probability field of cell-surface motor occupancy by centrosomal microtubules (under an assumption of stoichiometric binding), and free boundaries of unattached, growing microtubules. We show how cell shape affects the stability of centering of the aster, and its transition to oscillations with increasing motor number. Seeking to understand observations in single-cell nematode embryos, we use highly accurate simulations to examine the nonlinear structures of the bifurcations, and demonstrate the importance of binding domain overlap to interpreting genetic perturbation experiments. We find a generally rich dynamical landscape, dependent upon cell shape, such as internal constant-velocity equatorial orbits of asters that can be seen as traveling wave solutions. Finally, we study the interactions of multiple asters which we demonstrate an effective mutual repulsion due to their competition for surface force generators. We find, amazingly, that centrosomes can relax onto the vertices of platonic and non-platonic solids, very closely mirroring the results of the classical Thomson problem for energy-minimizing configurations of electrons constrained to a sphere and interacting via repulsive Coulomb potentials. Our findings both explain experimental observations, providing insights into the mechanisms governing spindle positioning and cell division dynamics, and show the possibility of new nonlinear phenomena in cell biology. 
\end{abstract}

\maketitle

\section{Introduction}

The centrosome, a micron-scale organelle \cite{decker2011limiting}, is the primary microtubule organizing center in animal cells and plays a central role in cellular processes such as division, polarization, and intracellular organization and transport \cite{bornens2012centrosome}. Microtubules (MTs), stiff polar biopolymers having a persistence length on the order of millimeters \cite{janson2004bending}, nucleate from the centrosome with their plus-ends growing outwards and their minus-ends tethered to the centrosome. A centrosome and its radially oriented MTs form the characteristic centrosomal aster. 

\begin{figure}[t]
\includegraphics[keepaspectratio=true]{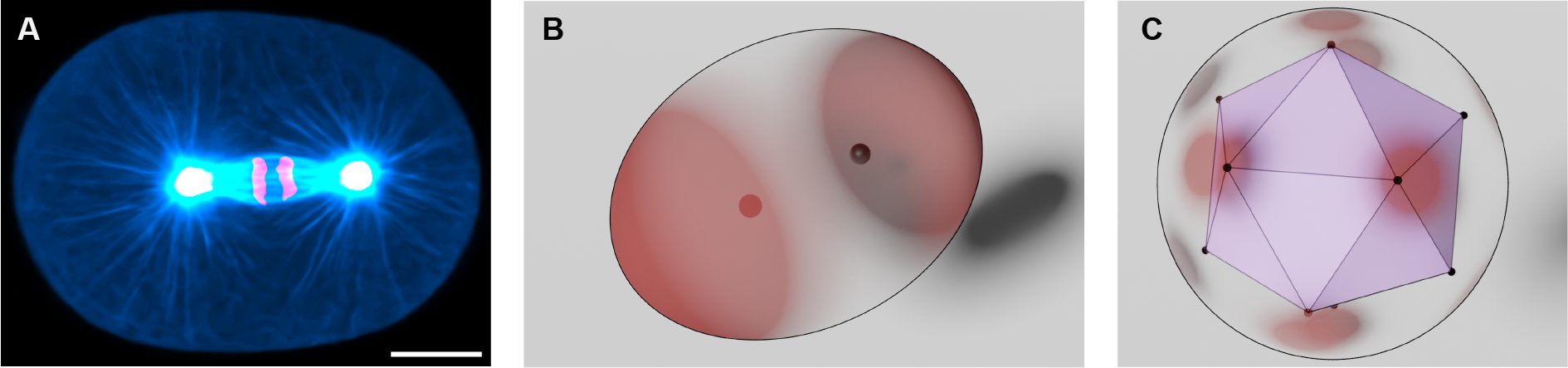}
\caption{Mitotic spindle and results from our stoichiometric model. (A) First mitotic spindle in \textit{C. elegans} embryo. Blue shows the microtubules, pink indicates the chromosomes, and white represents the centrosomes. Scale bar, $10\mu$m. See also Supplementary Video 1. (B) Equilibrium configuration of two centrosomes (black circles) inside a prolate cell. Color-field represents the probability that a CFG is bound to a MT. (C) Equilibrium configuration of twelve centrosomes that arranged into the vertices of an icosahedron embedded in a sphere.}
\label{Fig1_1}
\end{figure}

An important biophysical aspect of centrosomal asters is their ability to exert and transmit forces. For example, during cell division the mitotic spindle, a bipolar structure primarily composed of transitory MTs and associated proteins, forms near the cell center with a centrosome, and its aster, at each pole (Fig.~\ref{Fig1_1}A and Supplementary Video 1). Accurate spindle positioning, largely mediated by the centrosomal asters, is crucial for precise genome and organelle segregation, while errors in centrosome and spindle positioning can prove fatal for cells. Various forces upon centrosomal MTs, including MT polymerization-driven pushing forces against the cell cortex, pulling forces from motor proteins carrying payloads along MTs, pulling forces from cortically anchored motor proteins, and forces from MT friction with the cell wall, have been proposed to drive spindle positioning within cells~\cite{tran2001mechanism, grill2003distribution, grill2005theory, howard2006elastic, CampasSens2006_PRL, thery2007experimental, kozlowski2007cortical, hara2009cell, zhu2010finding, shinar2011model, laan2012cortical, pavin2012positioning, ma2014general, garzon2016force, pecreaux2016mitotic, letort2016centrosome, coffman2016asymmetric, howard2017physical, tanimoto2018physical, farhadifar2020stoichiometric}. While the exact force mechanism behind spindle positioning remains an open question, even for many model organisms, molecular and biophysical perturbation experiments across diverse cell types, including yeast~\cite{carminati1997microtubules}, \textit{C.~elegans}~\cite{grill2003distribution}, and human cells~\cite{kotak2012cortical}, have shown the dominance of pulling forces exerted by cortically anchored motor proteins, such as the minus-end directed motor dynein. 
%For example, in the early {\it C. elegans} embryo, the asymmetry in the MT nucleation rate between the two centrosomes before the entry into mitosis \cite{coffman2016asymmetric} may persist into metaphase and affect the spindle positioning.
A centrosome typically nucleates hundreds of MTs per second, each with a short lifespan of tens of seconds, collectively forming a dynamic aster comprising thousands of MTs. At any given moment a subset of these MTs, typically a few hundred, interfaces with the cortex, giving these MTs the potential to engage with dynein motors anchored there. In this scenario, an impinging MT binds to an anchored dynein motor, which then walks towards the MT's minus-end and so exerts a pulling force upon it~\cite{gusnowski2011visualization}. The dynein motor, together with its anchoring protein complex, is referred to as a cortical force-generators, or CFG. The interaction between CFGs and MTs is transient, perhaps from the motor's detachment from either cortex or microtubule, or, possibly, the disassembly of the MT itself, but collectively generates the tens of piconewton forces involved in positioning the centrosome, and other structures such as spindles to which it is connected, within the cell \cite{srayko2005identification, redemann2017c}. The precise role of these pulling forces in spindle positioning remains the subject of ongoing theoretical and experimental inquiry.

In prior work, we developed versions of the so-called stoichiometric model, or S-model, for how mobile centrosomal asters interact with CFGs and through them studied aspects of spindle dynamics~\cite{farhadifar2020stoichiometric, WuEtAl2023}. The S-model framework is based exclusively on the pulling forces exerted by CFGs upon MTs and, through the transience of their interactions, deals naturally with the cell geometry through which a centrosomal aster moves. It is derived from the fundamental biophysics of MT dynamics and well-established biochemistry of molecular motors. Central is a stoichiometric interaction between CFGs and MTs: a CFG can bind to only one MT at a time. Until detachment, the bound MT experiences a pulling force, which is transmitted to the centrosome. In \cite{farhadifar2020stoichiometric} we developed an S-model with discrete CFGs assuming a quasi-static balance of MT cortical impingement rate and motor-MT detachment. For {\it C. elegans} single-cell embryos this model quantitatively explained the dynamics of spindle positioning and elongation, and final length scaling with cell size, while accounting for their variations within species and across nematode species spanning over 100 million years of evolution. In~\cite{WuEtAl2023}, a combination of laser scission of MTs, cytoplasmic flow reconstructions, computational fluid dynamics, and an S-model, were used to demonstrate the predominance of cortical pulling forces in the dynamics of pronuclei and of the spindle in all stages leading up to the first cell division {\it C. elegans} embryo. There the S-model was extended to include the dynamics of MT-motor interactions, and the discrete CFGs replaced with a continuous surface distribution of occupancy probability. The S-model consists of an ordinary differential equation (ODE) for centrosome position driven by a surface integral of motor forces weighted by an evolving probability field of MT-motor attachment. Within a spherical cell under symmetry assumptions this S-model was mathematically analyzed to show, among other things, that stable positioning could transition to oscillations via a Hopf bifurcation, at motor densities and at oscillation frequencies consistent with observations of spindles during the metaphase to anaphase transition.

Given the ubiquity and importance of centrosomal asters to cellular dynamics, here we seek to understand this class of models more comprehensively. In \S~\ref{sec:CG1}, we develop an elaborated S-model, which, again, takes the form of an overdamped dynamics for centrosome position driven by a surface integral of coarse-grained MT-oriented pulling forces from CFGs, weighted by the probability for a CFG to be occupied by an MT. The probability evolves as a balance of binding by impinging MTs and unbinding. Unlike our earlier models, here we include the effect of random overlap in CFG binding domains. We show that the S-model has a natural energy-like quantity that evolves through a balance of input power from MT impingement and binding, and dissipation from drag and unbinding. In \S~\ref{sec:bifurcations} we study the centering of single centrosomes within spheroidal cells; linear analysis for a centrosomal aster  predicts stable centering which, with increasing motor density, loses its stability to oscillation via a Hopf bifurcation as suggested by experiment. We find that in the stably centered case, the elaborated S-model accounts for force-displacement experiments using genetic perturbations. Using highly accurate numerical methods we investigate the nonlinear structures of the Hopf bifurcation, seeking and finding supercriticality as is inferred from experiments probing the transition to spindle oscillation. In \S~\ref{EqOrbsThomson} we show that there is a rich variety of possible dynamics. We find and construct internal equatorial orbits of asters that arise from a symmetry breaking instability. We investigate how multiple centrosomes interact. Within the S-model, centrosomes compete for force-generators, with that competition creating an effective repulsion between them. Consequently, for two centrosomes we can observe relaxation to well-separated positions, as in Fig.~\ref{Fig1_1}B, reminiscent of the positioned mitotic spindle, as in Fig.~\ref{Fig1_1}A. This competition plays out beautifully when simulating larger numbers where asters move to the vertices of platonic (see Fig.~\ref{Fig1_1}C for 12 centrosomes relaxing to an icosahedron) and non-platonic solids in a manner remarkably similar what is found in J.J. Thomson's classical problem of electrons on a sphere interacting by a Coulomb potential \cite{tomson1904structure}.

% Finally, in a study reminiscent of J.J. Thomson's problem of electrons moving on the surface of a sphere, we simulate the dynamics of multiple asters moving within a spherical cell and find, for example, their relaxation to the vertices of platonic solids.

\section{A generalized formulation of the stoichiometric model\label{sec:CG1}}

\begin{figure}[t]
\includegraphics[keepaspectratio=true]{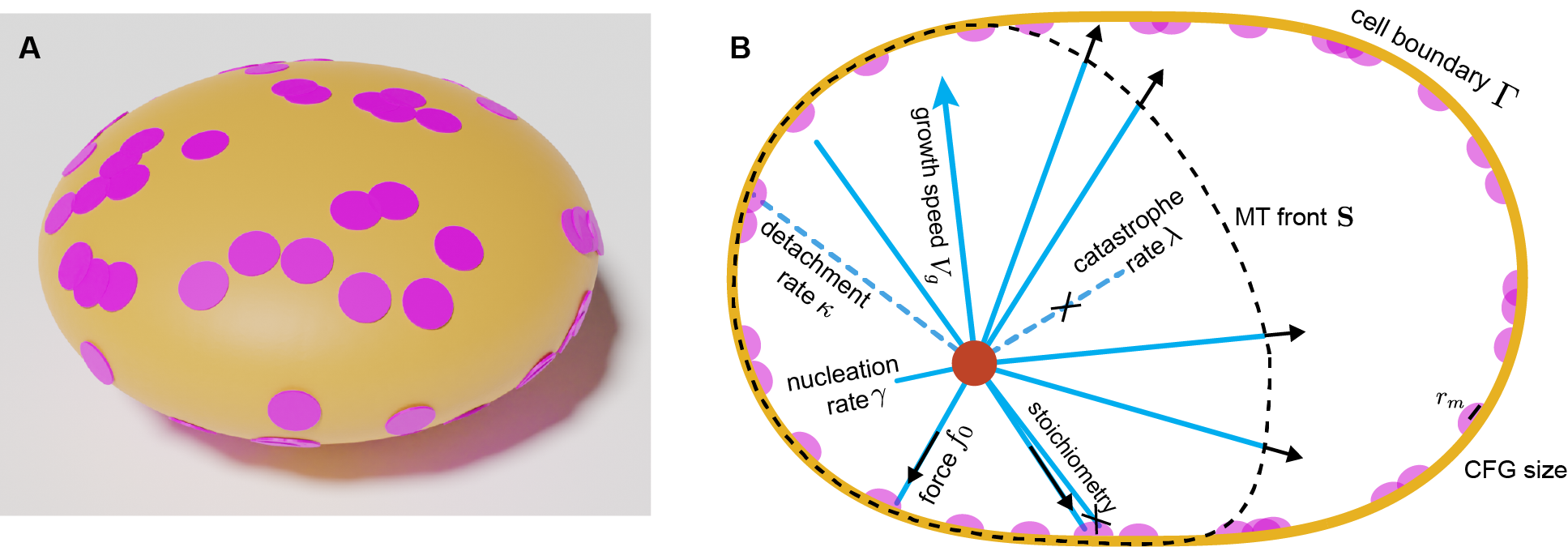}
\caption{Randomly distributed overlapped force generators on the cell surface and schematics of the stoichiometric model. (A) A view of the cell periphery (yellow) with randomly placed CFGs (pink circular disks). (B) Schematics of the S-model. Microtubules (MTs, blue) nucleate from the centrosome (red) with rate $\gamma$, grow with speed $V_g$, and undergo catastrophe with rate $\lambda$. The dashed line indicates the microtubule front ($\mathbf{S}$). Cortical force generators of size $r_m$ (CFGs, pink) are distributed on the cell surface $\Gamma$. The interaction of CGFs and MTs is stoichiometric - a CFG can bind only one MT at a time. Once bound, the CFG exerts a pulling force of magnitude $f_0$ along the MT to the centrosome. CFGs detach from MTs with rate $\kappa$ and become unbound.}
\label{Fig1_2}
\end{figure}

Building and expanding upon our prior studies \cite{farhadifar2020stoichiometric, WuEtAl2023}, we consider the cell periphery to be populated by CFGs in random positions, without considering an exclusion principle (Fig.~\ref{Fig1_2}A), and MTs as straight polymers nucleating with spherical uniformity at rate $\gamma$ from a centrosome at location $\mathbf{X}(t)$; see Fig.~\ref{Fig1_2}B. This is consistent with the relative rigidity of MTs and the observation that end-pulling forces create an extensional stress in the attached MTs~\cite{NRzs2017,NRNS2017}. With their minus-ends fixed at the centrosome, model MTs grow by polymerization at their plus-ends at a constant speed $V_g$, and undergo catastrophe with rate $\lambda$; see Fig.~\ref{Fig1_2}B.

An MT that reaches the cell surface $\Gamma$ has two possible fates: either it binds to an unoccupied cortical force-generator (CFG) anchored there or undergoes catastrophe and disassemble. We take a stoichiometric interaction between CFGs and MTs: a CFG can only be occupied by one MT at a time. Once a CFG is bound to an MT, it exerts a constant pulling force of magnitude $f_0$ on the centrosome toward the CFG; see Fig.~\ref{Fig1_2}B. Bound CFGs will also detach from their MTs with a rate $\kappa$ and become unoccupied. Neglecting inertia, the dynamics of the centrosome and its MT array are governed by a balance between pulling forces from the CFGs and the viscous drag forces from the cytoplasm on the centrosome and its attached MTs. 

Consider a centrosome at position $\mathbf{X}(t)$, hence with velocity $\dot{\mathbf{X}}(t)$, and at distance $D(\mathbf{Y},t)=|{\bf Y}-{\bf X}(t)|$ from points $\mathbf{Y}\in\Gamma$. We define two unit vectors: the outward normal $\hat{\mathbf{n}}(\mathbf{Y})$ to $\Gamma$, and $\hat{\boldsymbol{\xi}}(\mathbf{Y};{\bf X})=({\bf Y}-{\bf X}(t))/D({\bf Y},t)$, the direction along which forces are exerted by CFGs.  

We assume a total number $M$ of CFGs, whose centers are independent and identically distributed with a probability density $\rho$ (i.e. $\int_{\Gamma} \rho({\bf Y}) dA = 1$). Note that independence implies that there is no mutual exclusion between CFGs, hence there can be overlap between the binding domains of different CFGs. Further, we assume that $\rho$ smoothly varies across $\Gamma$. Note, for a globally uniform distribution of CFGs $\rho(\mathbf{Y})\equiv 1/|\Gamma|$ where $|\Gamma|$ is the surface area of $\Gamma$.

For sufficiently large $M$, we can apply the law of large numbers, so up to statistical fluctuations, we can express the total pulling force as an integral, and assume the centrosome moves by the overdamped dynamics
\begin{align}
\label{eq:Zdot_3d}
    \eta \dot{\bf X} = M f_0\int_{\Gamma} P({\bf Y},t)\hat{\boldsymbol{\xi}}({\bf Y};{\bf X})\,\rho({\bf Y})dA + \mathbf{F}_{ext}, 
\end{align}
where $P({\bf Y}, t)$ is the probability that a CFG centered at ${\bf Y}$ is occupied by an MT, and $\eta$ is a drag coefficient. Note that we have allowed for the action of an external force $\mathbf{F}_{ext}$ upon the centrosome so as to mimic particular experiments. 

Introducing $P({\bf Y}, t)$ instead of the discrete states of a CFG being occupied (bound) or unoccupied (unbound), we have effectively introduced a coarse-graining in time, allowing us to write a forward-Kolmogorov equation. However, to properly do this, we need to do a spatial coarse-graining of the CFGs. In particular, we assume that each CFG can bind over a discal area of radius $r_m$ (with area $a_m=\pi r_m^2$) and that $P({\bf Y}, t)$ has a slow variation through the surface. Considering stoichiometry of the CFGs, it follows that the evolution equation for the occupancy probability $P$ is
\begin{align}
\label{eq:dPdt_overlap}
    \frac{\partial P({\bf Y},t)}{\partial t} 
    &= \Omega(\mathbf{Y},t){\cal I}(P, {\bf Y}) - \kappa P({\bf Y},t)~\mbox{with} \\
    {\cal I}(P,\mathbf{Y})&= \frac{1-e^{-a_mM\rho(\mathbf{Y})
    (1-P({\bf Y},t))}}{a_mM\rho(\mathbf{Y})}.
    \nonumber
\end{align}
Here $\Omega(\mathbf{Y},t)$ is the MT impingement rate on $\Gamma$, or the number of unbound MTs per unit time that hit the area covered by a CFG centered at $\mathbf{Y}$, and ${\cal I}(P, {\bf Y})$ is the probability that an MT succesfully binds to the CFG whose binding domain is centered at ${\bf Y}$ (for a complete derivation, see \S~\ref{subsec:overlap_model}).

We can understand ${\cal I}(P, {\bf Y})$ further by examining its form. First, note that $0 \leq {\cal I} \leq 1$ and $\partial_P{\cal I} <0$ for $P\in [0,1]$. The first bound is trivial, as ${\cal I}$ is a probability, and the second one is due to the assumption of stoichiometry (the larger the $P$, the less likely a new MT binds to the CFG centered at $\bf Y$). Secondly, $a_mM\rho(\mathbf{Y})$ approximates the number of CFGs covering the point ${\bf Y}$, and thus is a measure of competition between them. For  $a_mM\rho \ll 1$ (or $1-P \ll 1$), i.e. negligible overlap (or CFGs are close to complete saturation), we have ${\cal I} \rightarrow 1-P$, recovering the {\it independent CFG model}, as used in  \cite{farhadifar2020stoichiometric, WuEtAl2023}:
\begin{align*}
    \frac{\partial P({\bf Y},t)}{\partial t} 
    &= \Omega(\mathbf{Y},t)\left(1-P\left({\bf Y},t\right)\right) - \kappa P({\bf Y},t),
\end{align*}
where the competition between CFGs becomes irrelevant either due to the low level of overlap or because CFGs are almost always occupied and do not locally compete with each other. For $(1-P) a_mM\rho \gg 1$, ${\cal I} \approx \frac{1}{a_mM\rho}$,  signifying that an MT has an equal chance of binding to any of the CFGs covering the same area, making the competition between CFGs the most relevant. Lastly, the numerator $1-e^{-a_mM\rho(\mathbf{Y})(1-P({\bf Y},t))}$ is the probability that at least one of the CFGs at ${\bf Y}$ is unbound (\S~\ref{subsec:overlap_model}). The term $\frac{1}{a_mM\rho(\mathbf{Y})}$ comes from the fact that since all CFGs are indistinguishable, the incoming MT will choose one of them randomly. 

Following \cite{WuEtAl2023} we approximate the impingement rate $\Omega$ by
\begin{align}
\label{eq:impingement_rate_3d}
    \Omega(\mathbf{Y},t) = \phi({\bf S}) 
    \left[{\bf V}_S\cdot\hat{\bf n} \right]_+
    \chi\left(\frac{r_m}{D}\right)
    \left(\frac{\gamma}{V_g} e^{-D/l_c}\right).
\end{align}
In this expression, the first term $\phi({\bf S})$ is the indicator function specifying whether the MT front (Fig.~\ref{Fig1_2}B) is in contact with $\Gamma$ (for which $\phi({\bf S})=1$), or not (for which $\phi({\bf S})=0$).  The second term contains $\mathbf{V}_S(\mathbf{Y},t)=\dot{\mathbf{X}}+V_g\hat{\boldsymbol{\xi}}(\mathbf{Y}; {\bf X})$, the velocity of an MT plus-end with a direction $\hat{\boldsymbol{\xi}}(\mathbf{Y};{\bf X})$, and $\left[{\bf V}_S\cdot\hat{\bf n} \right]_+$ gives the rectified normal component of that velocity at $\Gamma$; If ${\bf V}_S\cdot\hat{\bf n}<0$ the centrosome is moving away from the surface faster than MTs grow, and thus the impingement rate is zero.  The third term $\chi\left( r_m/D \right) = \frac{1}{2}\left(1-\frac{1}{\sqrt{1+(r_m/D)^2}}\right)$ estimates the fraction of centrosomally-nucleated, unbound MTs at $\Gamma$ available for binding to a CFG. Finally, $\frac{\gamma}{V_g} e^{-D/l_c}$ is the equilibrium astral MT number per unit length reaching $\Gamma$ from the centrosome (at distance $D$), with $l_c=V_g/\lambda$ the characteristic length of MTs undergoing dynamic instability. We note that for time $t>W/V_g$ it is appropriate to use the steady state distribution because of linear hyperbolic nature of the associated Fokker-Plank equation for the MT length distribution~\cite{WuEtAl2023}.

Next, we consider the dynamics of the MT front (surface ${\bf S}$ in Fig.~\ref{Fig1_2}B), which incorporates the finite propagation speed of MT plus-ends and accounts for the detachment of the MT front due to the centrosome moving away from the cell boundary faster than MT growth. The full front is described by the equation
%%
%Put equation here
\begin{align}
\label{eq:dSdt}
    \frac{d {\bf S}}{dt} &= \phi({\bf S}) \left[-{\bf V}_S\cdot\hat {\bf n}\right]_+{\bf V}_S + \left(1-\phi({\bf S})\right){\bf V}_S\,,
\end{align}
Note that when the front is not located in the cell periphery, it translates with the centrosome expanding at velocity $V_g$, but when a point on the MT front is located on the cell periphery, it either stays on the periphery or moves with the velocity ${\bf V}_S$ inwards and away from the cell boundary.

The model, Eqs.~(\ref{eq:Zdot_3d}-\ref{eq:dSdt}), has seven parameters (see Table~\ref{tab:parameters}). We take the cell scale $W=\sqrt{|\Gamma|/4\pi}$ as a characteristic length scale, and $\tau=W/V_g$ (the time scale for an MT to grow from cell center to cell periphery) as a characteristic time scale. Other faster time scales are the inverse nucleation rate $1/\gamma$ (fastest) and the inverse detachment rate $1/\kappa$ (next fastest). The smallest length scale is $r_m$. From these we obtain five dimensionless parameters: dimensionless nucleation rate $\bar{\gamma}=\tau \gamma$, dimensionless detachment rate $\bar{\kappa}=\tau \kappa$, dimensionless MT length $\bar{l}_c=l_c/W$, dimensionless CFG size $\bar{r}_m=r_m/W$, and dimensionless force $\bar{f}_0=f_0/\eta V_g$. 

\medskip
\noindent{\bf Comments:}

\noindent{\bf i.}
The spatial coupling of $P(\mathbf{Y},t)$ occurs only through the integral coupling provided by Eq.~(\ref{eq:Zdot_3d}), and not through the explicit $P$ dynamics, Eq.~(\ref{eq:dPdt_overlap}), which is entirely local in $\mathbf{Y}$. Eqs.~(\ref{eq:Zdot_3d},\ref{eq:dPdt_overlap}) are a countably infinite set of ODEs when $P$ is expanded in a countable and complete set of surface basis functions (such as spherical harmonics).

\medskip
\noindent{\bf ii.}
The probability ${\cal I}(P)$ can take different forms, depending upon modeling assumptions. As discussed above, in the limit $a_mM/|\Gamma|\ll 1$, where the overlap between CFGs is negligible, ${\cal I}(P)$ in Eq.~(\ref{eq:dPdt_overlap}) simplifies to ${\cal I} = 1-P$, and the occupancy probability satisfies the independent CFG model. This was the form used in \cite{WuEtAl2023}, and earlier in discrete form by \cite{farhadifar2020stoichiometric}. In recent work in human spindles, we consider localized clusters of $N$ CFGs that are independent between each other, which yields an interaction in the form ${\cal I} = \frac{1-P^N}{N}$. 
%{\color{red} Going before out work, a version of stoichiometry has also been proposed to model chromosome oscillation in human cells ~\cite{CampasSens2006_PRL}, where a bath of $N$ chromokinesins are interacting with MT. In the context of our model, the corresponding interaction term used was ${\cal I} = 1-\frac{P}{N}$ }.  

\medskip
\noindent{\bf iii.}
Formally, one can write the RHS of Eq.~(\ref{eq:Zdot_3d}) as the gradient of an energy
\begin{align}
\label{eq:gradient_flow}
    \eta\dot{\bf X} &= -\nabla_{{\bf X}} {\cal E},
\end{align}
where the energy ${\cal E}$ is defined as 
\begin{align}
\label{eq:Energy}
{\cal E}[P;\mathbf{X}] &= M f_0\int_{\Gamma} P({\bf Y},t)D({\bf Y},t)\rho({\bf Y}) dA\geq 0,
\end{align}
recalling that $D(\mathbf{Y},t)=\left | \mathbf{Y}-\mathbf{X}(t) \right |$. This energy achieves its minimum of zero only for $P\equiv 0$, that is when there are no bound microtubules (and no velocity). Were $P$ independent of time, which it is typically not, the energy would change only with the centrosome position and so ${\cal E}$ would decay in time due to Eq.~(\ref{eq:gradient_flow}). Instead, using Eqs.~(\ref{eq:Zdot_3d},\ref{eq:dPdt_overlap}), ${\cal E}$ evolves as
\begin{align}
\label{eq:Edot}
    \dot{\cal E} &= M f_0 \int_{\Gamma}\Omega({\bf Y},t){\cal I}(P) D({\bf Y},t) \rho({\bf Y})dA 
    - \left(\kappa {\cal E} + \eta \left|\dot{\bf X}\right|^2\right)
    =\cal{P}-\cal{D},
\end{align}
where ${\cal P}\geq 0$ is the input power due to microtubule impingement and binding, and ${\cal D}\geq 0$ are the dissipations arising from microtubule detachment from CFGs and the viscous drag of aster motion. Hence, if ${\cal E}$ were in a nonzero steady-state this would reflect a balance of positive input power and positive dissipation.

\medskip
\noindent{\bf iv.}
The S-model generalizes naturally to the dynamics of $N$ centrosomes, with each ($i^{th}$) centrosome at its position ${\bf X}^i(t)$ having its surface probability $P^i({\bf Y},t)$ of CFG binding. In this generalization, all centrosomal asters compete for the CFGs, which is a fixed resource, through their impingement rates $\Omega^i({\bf Y})$. This competition is found in the dynamics of $P^i({\bf Y},t)$:
\begin{equation}
    \label{eq:multicentrosomes_p}
    \frac{\partial P^i({\bf Y},t)}{\partial t} =   \Omega^i(\mathbf{Y},t){\cal I}(P)-\kappa P^i({\bf Y},t),
\end{equation}
where $P=\sum\limits^N\limits_{i=1}P^i$ is the total CFG occupancy probability. The total probability $P$ then satisfies
\begin{align}
\label{eq:multicentrosomes_P}
    \frac{\partial P({\bf Y},t)}{\partial t} &=   \left(\sum\limits^N\limits_{i=1}\Omega^i(\mathbf{Y},t)\right){\cal I}\left(P\right) - \kappa P({\bf Y},t).
\end{align}
The earlier S-model of \cite{farhadifar2020stoichiometric}, using $N=2$, was used there to study the effects of motor competition on the dynamics and steady-states of two asters, as a way of understanding spindle length selection. Likewise, there is a multi-centrosome analog to Eqs.~(\ref{eq:Energy},\ref{eq:Edot}) for a total energy ${\cal E}=\sum\limits^N\limits_{i=1}{\cal E}^i$.

\subsection{Simulation Methods\label{subsec:simulation_methods}}

To numerically evolve ${\bf X}$ and $P$ via Eqs.~(\ref{eq:Zdot_3d}) and (\ref{eq:dPdt_overlap}), respectively, we first compute the surface integral of $P$ against $\hat{\boldsymbol{\xi}}$ in Eq.~(\ref{eq:Zdot_3d}), given ${\bf X}$ and $P$. Two different quadrature schemes are used to evaluate the surface integral: For the simulations in spheres examining oscillations near the Hopf bifurcation (\S~\ref{subsec:NonlinearStructureHopf}), $P$ is expanded in Legendre polynomials (\S~\ref{subsec:spectral_method}) and a high-order Gauss-Legendre quadrature is used to evaluate the integral. Otherwise, we map the cell surface to the unit sphere and use a more general patch-based high-order quadrature (\S~\ref{subsec:Methods_arbitrary_geometry}). The centrosome position is updated using the explicit Euler method with a time-step $\Delta t$ which also yields $\dot{\bf X}$ at the current time-step. Using this centrosome velocity and accounting for MT growth we then update the position of the microtubule front $\bf{S}$ position via an Euler step, whether $\mathbf{S}$ was in contact with $\Gamma$ or not. There are two cases. If the updated $\mathbf{S}$ lies within the cell, i.e. remains or becomes detached,  then the impingement rate imposed upon the line-of-sight CFGs (i.e. from the centrosome) is zero. If the front has moved outside of the cell then the front is projected back to $\Gamma$ and $\Omega$ is calculated for updating $P$ in those regions. For stability, we use a mixed explicit/implicit Euler scheme. The term ${\cal I}(P)$ is treated explicitly, while the damping term $-\kappa P$ is treated implicitly. This yields an explicit expression for $P$ at the next time step (Eq.~(\ref{eq:Pnp1}) in \S~\ref{subsec:TrackingMT}). The local error in time-stepping is found to be quadratic in $\Delta t$, and the global error scales linearly with $\Delta t$, as expected.

\section{Bifurcations and their nonlinear structures \label{sec:bifurcations}}

\begin{figure}[t]
\includegraphics[keepaspectratio=true]{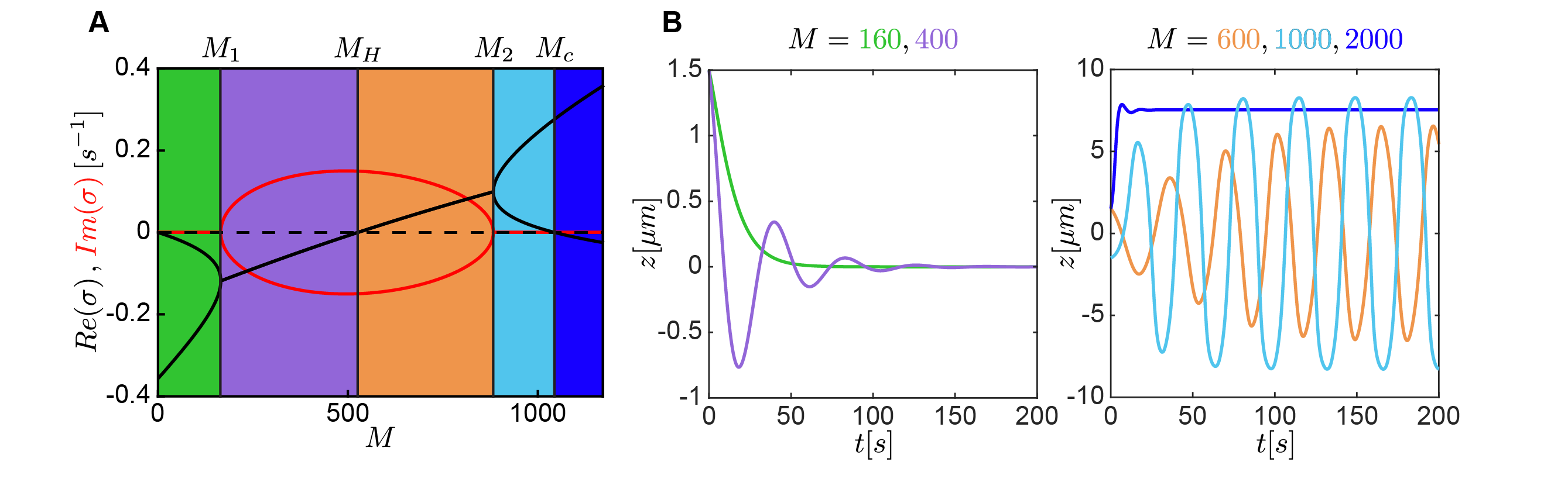}
\caption{Linear analysis of the S-model and on-axis centrosome dynamics in the sphere. (A) Linear solutions of the eigenvalues as a function of the bifurcation parameter $M$. Values of $M$ where a change in the centrosome dynamics occurs, are indicated. For $M<M_1=166$, the centrosome stably positions at the cell center. For $M_1<M<M_H=526$, the initially displaced centrosome moves to the center with damped oscillation. For $M_H<M<M_2=883$, the centrosome oscillates around the center with a growing amplitude that plateaus.  For $M>M_2$, the center is unstable and the centrosome decenters without oscillation. For $M>M_c$, the coefficient of the force-displacement changes sign. (B) Examples of centrosome trajectories for various values of $M$ as a function of time are shown.}
\label{Fig2}
\end{figure}
 
We first examine the linear behavior of the generalized S-model under small perturbations. For ease of calculation we assume that cell shapes are convex, axisymmetric about the $z$-axis, and symmetric across the $xy$-plane. Thus the origin, or the cell center $X_0$, is naturally a fixed point of centrosome position. It has an associated steady-state occupancy probability $P=P_0(\mathbf{Y})$ satisfying
\begin{equation}
\int_{\Gamma}P_0(\mathbf{Y})\hat{\mathbf{Y}} \,\rho(\mathbf{Y})dA_Y
=\mathbf{0},
~~\mbox{and}~~
P_0({\bf Y})=\Omega_0({\bf Y}){\cal I}\left(P_0\right)/\kappa
\label{eq:SteadyState}
\end{equation} 
where
\begin{equation}
\Omega_0( {\mathbf Y} ) = \hat{\mathbf{Y}}\cdot\hat{\mathbf{n}}({\bf Y})~
    \chi\left( \frac{r_m}{|\mathbf{Y}|}\right)e^{-\frac{| \mathbf{Y} |}{l_c}}\gamma
    \label{eq:R0}  
\end{equation}
and $\hat{\mathbf{Y}}=\mathbf{Y}/|\mathbf{Y}|$. Note that, under these conditions, cell convexity implies that the MT front is in contact with the cell surface at all points ($\phi(\mathbf{S})\equiv 1$). Also, for a general cell shape the nonlinear equation for $P_0$ must typically be solved numerically.

Consider perturbations around the fixed point, $\mathbf{X}(t)=\varepsilon\mathbf{x}(t)$ ($\dot{\mathbf{X}}=\varepsilon\dot{\mathbf{x}}$) and $P({\bf Y}, t)=P_0({\bf Y})+\varepsilon P_1({\bf Y}, t)$, likewise with a small external force $\mathbf{F}_{ext}=\varepsilon\mathbf{f}_{ext}$, where $\varepsilon\ll 1$. We can also assume $|\dot{\mathbf{X}}|\ll V_g$, ensuring continuous contact between the MT front and the cell surface. Therefore, we can disregard the MT front dynamics in Eq.~(\ref{eq:dSdt}). Expanding in $\varepsilon$ and truncating at linear order, from Eqs.~(\ref{eq:Zdot_3d},\ref{eq:dPdt_overlap}) we find the linearized equations of motion to be
\begin{align}
\label{eq:dX1dt_general}
      &\tilde{\eta} \dot{\bf x}(t)  = -  \int_{\Gamma} \frac{P_0({\bf Y})}{|\mathbf{Y}|} \left(  {\bf I} - \hat{\mathbf{Y}} \hat{\mathbf{Y}}\right) \rho({\bf Y}) dA_Y\cdot {\bf x}(t) 
     +  \int_{\Gamma} P_1({\bf Y},t)\hat{\mathbf{Y}} \rho({\bf Y}) dA_Y  + \tilde{\mathbf{f}}_{ext}.\\
     \label{eq:dP1dt_general}
       &  \frac{\partial P_1}{\partial t}({\bf Y}, t) 
    = \kappa P_0({\bf Y}) \, \hat{\mathbf{n}}({\bf Y}) \cdot \left(\frac{\dot{\bf x}(t)}{V_g}
    +  {\bf A}_1({\bf Y}) \cdot {\bf x}(t) \right) -\tilde \Omega_0({\bf Y}) P_1({\bf Y},t),
\end{align}
where $\tilde{\eta} = \frac{\eta}{M f_0} $ and $\tilde{\mathbf{f}}_{ext} = \frac{\mathbf{f}_{ext}}{Mf_0}$, and with
\begin{align}
    {\bf A}_1 ( {\mathbf Y}) &= -\frac{1}{\lvert \mathbf{Y} \rvert}   \left[  \mathbf{I}  - \left(1 + \frac{| \mathbf{Y} |}{l_c} + \frac{r_m}{| \mathbf{Y} |}
    \frac{\chi'\left(\frac{r_m}{| \mathbf{Y} |}\right)}
    {\chi\left(\frac{r_m}{| \mathbf{Y} |}\right)}
    \right)  
    \hat{\mathbf{Y}} \hat{\mathbf{Y}}  \right], \\
    \label{eq:R1} 
    \tilde \Omega_0({\bf Y}) &=\kappa+\Omega_0({\bf Y})e^{-a_mM\rho\left({\bf Y}\right)(1-P_0({\bf Y}))} > 0 .
\end{align}
The matrix ${\bf A}_1$ is symmetric and nonsingular.

We first perform linear stability analysis of Eqs.~(\ref{eq:dX1dt_general},\ref{eq:dP1dt_general}) in the absence of an external force (${\mathbf f}_{ext} = {\mathbf 0}$) and, second, compute the linear force-displacement relation for a stationary centrosome ($\dot{{\mathbf x}} = {\mathbf 0}$). Analytical results are derived for spherical geometry, and semi-analytic extensions are obtained for spheroidal geometries. For simplicity, we now assume the CFG distribution is uniform ($\rho\equiv 1/|\Gamma|$). 

\subsection{Linearized dynamics in a spherical cell
\label{subsec:stability_in_sphere}}

Consider a spherical cell of radius $W$, for which the equations above now simplifies drastically. In this case $\hat{\mathbf{Y}}=\hat{\mathbf{n}}$, $|\mathbf{Y}|=W$, and $\Omega_0$, $\tilde \Omega_0$, and $P_0$ are all constants. In brief sketch,
we use the fact that $\hat{ \bf n} \cdot \mathbf{A}_1\parallel\hat{\mathbf{Y}}$ and the identity $\int_{\Gamma} \hat{\mathbf{Y}}\hat{\mathbf{Y}} dA_Y = \frac{|\Gamma|}{3} {\bf I}$ to show that Eqs.~(\ref{eq:dX1dt_general},\ref{eq:dP1dt_general}) take the form
\begin{align}
\dot{\mathbf{x}}=-a\mathbf{x}+b\int_\Gamma P_1\hat{\mathbf{Y}} dS_Y+\mathbf{g}_{ext}
~~\mbox{and}~~P_{1t}=\mathbf{V}(t)\cdot\hat{\mathbf{Y}}-\tilde \Omega_0 P_1,
\nonumber
\end{align}
where $\mathbf{V}=\alpha\mathbf{x}+\beta\dot{\mathbf{x}}$, and $a$, $b$, $\alpha$, and $\beta$ are all positive constants and ${\bf g}_{ext} = {\tilde{\bf f}}_{ext}/{\tilde \eta}$. We note that $\int_\Gamma P_1\hat{\mathbf{Y}} dS_Y$ is the projection of $P_1$ against the three
 spherical harmonics (the components of $\hat{\bf Y}$) with $l=1$ polar index. A linear combination of these three $l=1$ harmonics also emerges in the forcing term, $\mathbf{V}(t)\cdot\hat{\mathbf{Y}}$, in the evolution equation for $P_1$. Consequently this system can be decomposed by setting $P_1=\mathbf{E}\cdot\hat{\mathbf{Y}} + Q(\mathbf{Y},t)$ where $Q$ is the projection of $P_1$ onto the spherical harmonics with $l\neq 1$, that is, those orthogonal to $\hat{\mathbf{Y}}$. We then find
\begin{equation}
    \dot{\mathbf{x}}=-a\mathbf{x}+\frac{b|\Gamma|}{3}\mathbf{E}+\mathbf{g}_{ext}, ~~
    \dot{\mathbf{E}}=(\alpha\mathbf{x}+\beta\dot{\mathbf{x}})-\tilde \Omega_0\mathbf{E}, ~~\mbox{and}~~
    Q_t=-\tilde \Omega_0 Q
    \nonumber
\end{equation}
So, for a sphere, the linearized Eqs.~(\ref{eq:dX1dt_general},\ref{eq:dP1dt_general}) can be decomposed into {\bf (i)} a 6-dimensional set of ODEs (for $\mathbf{x}$ and $\mathbf{E}$) in which each spatial direction can be treated independently and identically (modulo the direction of the external force), and which couple positional dynamics $\mathbf{x}$ to the evolving $l=1$ components $\mathbf{E}$ of surface probability $P_1$; {\bf (ii)}, an infinite number of surface harmonics evolved in $Q$, all of which decay to zero at rate $\tilde \Omega_0$. We find a similar but somewhat more complex structure for the generalized spheroid case. This generalizes and clarifies the analysis of \cite{WuEtAl2023} who assumed an axisymmetric $P_0$ and centrosome motion along the $z$-axis, and did not analyze the dynamics of the remainder modes in $Q$. 

The linearized dynamics of the system can now be understood through two scalar ODEs (say, by projecting along the direction of the external force). From these we can obtain a second-order ODE for the linear dynamics:
\begin{align}
\label{eq:linear_dynamics_sphere}
\tilde{\eta} \ddot{x} + \bigg( \tilde{\eta} \tilde{\Omega}_0 + \frac{P_0}{3 W} \left( 2 - \kappa\frac{W}{V_g} \right) \bigg) \dot{x} + 
\frac{P_0}{3 W} \left( 2\tilde{\Omega}_0 - \kappa \mathcal{B}  \right)x = \tilde{\Omega}_0 { \tilde{f} }_{ext} + \dot{ \tilde{f}}_{ext}\,,
\end{align}
where ${\cal B} = \left(W/l_c + \frac{r_m}{D}\frac{\chi'(\frac{r_m}{D})}{\chi(\frac{r_m}{D})}\right) = W/l_c + 2 + \mathcal{O}( (\frac{r_m}{D} )^2)$. To examine the linear stability of the fixed point we set $\tilde{f}_{ext}$ to zero, assume $x\sim e^{\sigma t}$, and determine the two eigenvalues $\sigma_{1,2}$. 

Turning to a specific case, consider the linearized dynamics for the parameter values in Table~\ref{tab:parameters} and Fig.~\ref{Fig2}: for $M< M_1 \, ( \sim 165)$, the two eigenvalues are real and negative, resulting in stable positioning of the centrosome at the cell center (Fig.~\ref{Fig2}A\&B). These two real eigenvalues collide at $M=M_1$, and a complex pair of eigenvalues with negative real parts emerges, giving damped oscillations of the centrosome when perturbed from the cell center for $M_H> M > M_1$. At $M=M_H \, (\sim 525)$ the real part crosses zero, becoming positive with nonzero imaginary part, that is, this is a Hopf bifurcation. The centrosome spontaneously oscillates with frequency $\omega_H=\sqrt{Mf_0P_0(2\tilde{\Omega}_0-\kappa\mathcal{B})/3W\eta}\sim 0.16 s^{-1}$. For $M_H< M < M_2\,( \sim 880)$, the real part of the eigenvalues is positive, resulting in an oscillatory motion around the center. For $M_2<M<M_c$, both eigenvalues are real and positive. For $M>M_c$, the eigenvalues are real with different signs. Thus, the S-model exhibits almost every allowable type of linear dynamics of a second order ODE as a function of CFG number $M$.

For illustration, Figs.~\ref{Fig2}B show examples of the nonlinear dynamics in these different regimes assuming axisymmetry of $P$ and motion strictly on the $z-$axis. The left panel shows strict relaxation and damped oscillations. The right panel shows on-axis oscillations and strict decentering $(M=2000)$ of the centrosome. The curves are color-coded by the intervals of $M$ in Fig.~\ref{Fig2}A.

\subsection{Force versus displacement \label{subsec:mechanics_centrosome_centering}}

\begin{figure}[t]
\includegraphics[keepaspectratio=true]{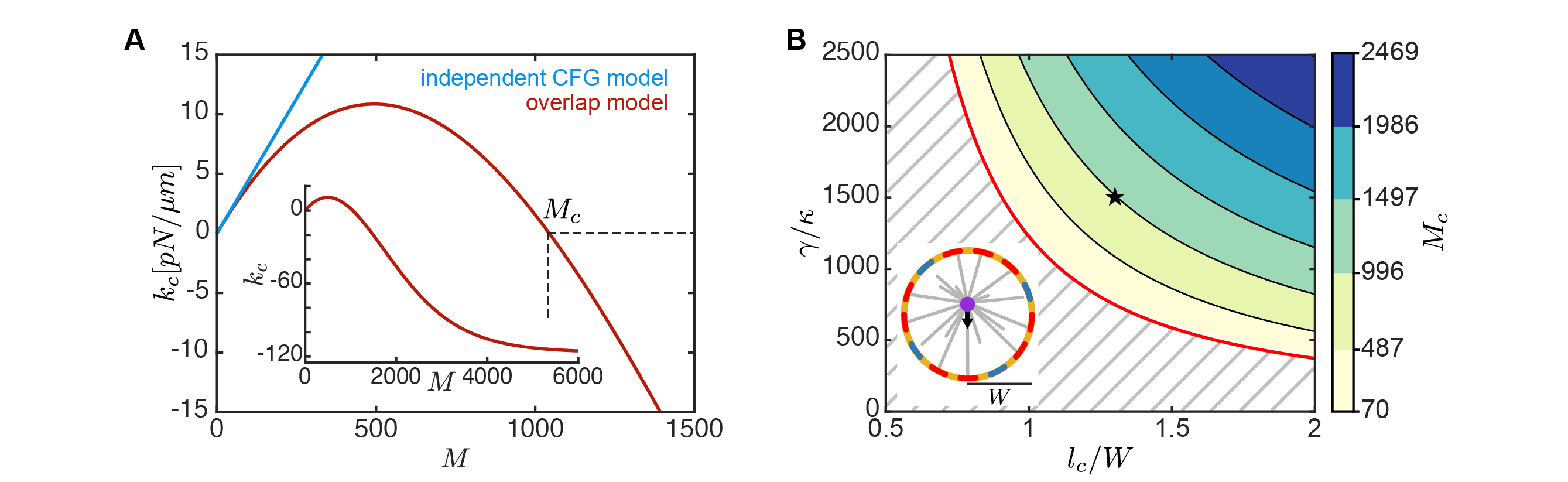}
\caption{Stable aster centering in the sphere.
(A) Force-displacement coefficient, $k_c$, as a function of CFG number $M$ for the independent CFG model (blue) and the overlap model (red) using parameter values in Table~\ref{tab:parameters}. At $M=M_c$, $k_c$ becomes negative, and the center becomes unstable. The inset shows $k_c$ for a larger range of $M$. (B) The value of $M_c$ as a function of dimensionless MT length ($l_c/W$) and $\gamma/\kappa$. In the dashed region, $M_c=0$ and the center is unstable for any range of $M$. The red line shows the analytical result for the stable to unstable transition. The inset shows the schematic of the simulation, where red/blue indicates bound/unbound CFGs. The star indicates parameters in Table~\ref{tab:parameters}.}
\label{Fig4}
\end{figure}

A number of experimental studies have examined the mechanics and stability of spindle centering by exerting a controlled external force on the spindle and measuring the resulting displacement~\cite{garzon2016force,tanimoto2018physical,anjur2023clustering}. For example, measurements in early \textit{C. elegans} embryos revealed that a micron displacement of the spindle from the cell center required forces of $\sim 10\;p\text{N}$~\cite{garzon2016force}. Further, it has also been shown that in mutant embryos, where motors are partially depleted, a higher force is required to displace the spindle~\cite{garzon2016force}. It was argued that this behavior arose from a combination of pulling forces from CFGs acting to decenter the centrosome, and pushing forces from MTs growing against the cell periphery acting as centering ones~\cite{garzon2016force}. Nonetheless, when forces upon the centrosome were probed via laser cutting~\cite{WuEtAl2023} no substantial contributions from pushing forces was found.

Having found stable centering with only pulling forces in the S-model, it is natural to enquire the magnitude of the centering force and its variation with model parameters. To do this we set $\ddot{x}=\dot{x}=\dot{f}_{ext}=0$ in Eq.~(\ref{eq:linear_dynamics_sphere}), and calculate the force-displacement coefficient, $k_c=f_{ext}/x$:
\begin{align}
\label{eq:kc_overlap_FGs}
k_c = \frac{M f_0 P_0}{3 W} \left(2 - \frac{\kappa}{\tilde{\Omega}_0}\mathcal{B} \right).
\end{align}
For $M=200$ and parameter values in Table~\ref{tab:parameters}, $k_c\sim 7 \;p\text{N}/\mu m$, which agrees in magnitude with measurements in \textit{C. elegans}~\cite{garzon2016force}. While $k_c$ is often referred to as a spring coefficient the centrosome of the S-model simply does not have a simple spring-dashpot response to an applied force. 

Previously, for the independent CFG S-model~\cite{WuEtAl2023}, we showed that $k_c$ increases linearly with $M$ (blue curve in Fig.~\ref{Fig4}A), in apparent disagreement with experimental observation. This follows directly from Eq.~(\ref{eq:kc_overlap_FGs}) by setting ${\tilde{\Omega}_0}$ to a constant independent of $M$, which is true for the independent CFG S-model, but not for the overlap model; see Eq.~(\ref{eq:R1}). Figure~\ref{Fig4}A (red) shows $k_c$ for the overlap S-model. For low $M$, $k_c$ increases monotonically with $M$, consistent with the independent CFG model, but is locally convex down and reaches a maximum of $k_c\sim 11 \;p\text{N}/\mu m$ around $M=495$. Beyond that, $k_c$ decreases monotonically with increasing $M$ (and eventually becomes negative for sufficiently large $M$). This region of increasing $k_c$ with decreasing $M$, due to the effect of overlap between CFGs, could be an explanation for experimental measurements in \textit{C. elegans}, where a decrease in motor number increases the centering stiffness \cite{garzon2016force}.

Our analysis shows that for a sufficiently large CFG number $M\ge M_c$, where one of the real eigenvalues is positive and the other is negative (Fig.~\ref{Fig2}A), $k_c$ is negative (Fig.~\ref{Fig4}A), and the sphere center turns to a saddle fixed point. From Eq.~(\ref{eq:kc_overlap_FGs}), it is straightforward to show that $k_c$ is positive only if $\frac{W}{l_c}e^{-W/l_c}\frac{\kappa}{\gamma}\le 2\chi(r_m/W)$, where two dimensionless parameters $l_c/W$ and $\gamma/\kappa$ characterize the transition from centering to decentering. We calculated the critical value of CFG number $M=M_c$, at which this transition occurs as a function of $l_c/W$ and $\gamma/\kappa$.  We found that for low $l_c/W$ and $\gamma/\kappa$, the center is always unstable (Fig.~\ref{Fig4}B). However, for larger $l_c/W$ and $\gamma/\kappa$, the center becomes stable for $M$ smaller than $M_H$, and the decentering transition occurs at higher $M_c$ (Fig.~\ref{Fig4}B). 

\subsection{Linear behavior in a spheroid \label{subsec:stability_in_spheroid}}

\begin{figure}[t]
\includegraphics[keepaspectratio=true]{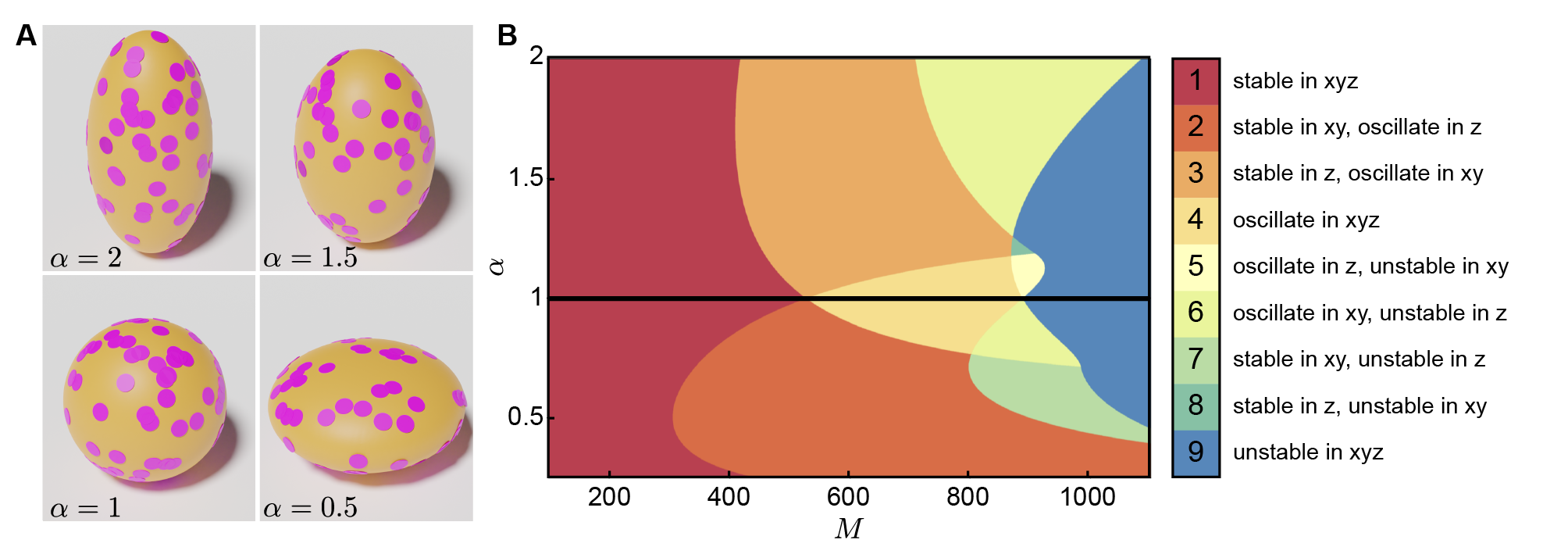}
\caption{Linear analysis of the S-model in spheroidal geometries.
(A) Four representative cell shapes parameterized by the shape factor $\alpha$ with the same surface area: prolate for $\alpha>1$, sphere for $\alpha=1$, and oblate for $\alpha < 1$. (B) Linear stability analysis of the centrosomal aster as a function of cell shape $\alpha$ and CFG number $M$ for parameter values in Table~\ref{tab:parameters}. The color code denotes the nine eigenmodes and associated centrosome dynamics, where ``stable" means that the real part of the eigenvalues is negative, ``oscillate" means that the eigenvalue has a positive real part and a non-zero imaginary part, and ``unstable" means that the real part is positive and the imaginary part is zero.}
\label{Fig3}
\end{figure}

In \S~\ref{subsec:stability_in_sphere}, we show that in spherical geometry, the linear dynamics of the centrosome is determined by two eigenvalues with an independent motion in each direction. While spheres, due to their symmetries, are amenable to analytical analysis, as the reader may suspect not all biological cells are spherical (e.g., see \textit{C. elegans} embryo in Fig.~\ref{Fig1_1}A and Supplementary Video 1). In this section, we study the linear behavior of the S-model for a class of axisymmetric spheroids centered at the origin and with the $z$-axis taken as the axis of symmetry. We show that only four eigenvalues determine the dynamics of the system- two are associated with the movement in the $xy$-plane, and the other two along the $z$-axis. 

To extend the stability analysis in \S~\ref{subsec:stability_in_sphere} to spheroids, we substitute the ansatz ${\bf x}(t) = {\bf v} e^{\sigma t}$ and $P_1({\bf Y, t}) = p({\bf Y}) e^{\sigma t}$ as solutions into the linearized Eqs.~(\ref{eq:dX1dt_general}, \ref{eq:dP1dt_general}) and obtain
\begin{align}\label{eq:spheroid}
\eta \sigma {\bf v} &= - \frac{M f_0}{|\Gamma|} \int_{\Gamma} \frac{P_0(\mathbf{Y})}{ |{\bf Y}|} \left(  {\bf  I} - \hat{\bf Y} \hat{\bf Y}^T \right)  dA_Y \cdot {\bf v} 
+ \frac{M f_0}{|\Gamma|} \int p({\bf Y}) \hat{{\bf Y}}  dA_Y \\
\label{eq:spheroid_p}
\sigma  p({\bf Y}) &= \kappa P_0(\mathbf{Y})  \left(  {\bf A}_1 \hat{\bf n} + \frac{ \sigma}{V_g} \hat{\bf n}  \right)\cdot {\bf v} - \tilde{\Omega}_0(\mathbf{Y})  p({\bf Y}).
\end{align}
Here, we focus on the eigenvalues that correspond to oscillatory/unstable centrosome motion, i.e., $\bf v \not = 0$ and $\sigma \not \in \mathbb{R}^{-}$. For a discussion of the stable spectrum see \S~\ref{subsec:stable_singular_spectrum}. By rearranging Eq.~(\ref{eq:spheroid_p}) we get
\begin{align}
p({\bf Y}) = \frac{\kappa P_0({\bf Y})}{\sigma + \tilde{\Omega}_0({\bf Y})} \hat{\bf n}^T \left(   {\bf A}_1  +   \frac{ \sigma }{V_g}  \right)  \cdot {\bf v},
\label{p_in_v}
\end{align}
which we substitute to Eq.~(\ref{eq:spheroid}) and obtain the nonlinear eigenvalue equation
\begin{align}
\eta \sigma {\bf v} = \frac{Mf_0}{|\Gamma|} \left[ - \int_{\Gamma} \frac{P_0}{ |{\bf Y}|} \left(  {\bf  I} - \hat{\bf Y} \hat{\bf Y}^T \right)  dA_{\bf Y} 
+ \int_{\Gamma} \kappa P_0 \frac{ V_g \hat{\bf Y} \hat{\bf n} ^T {\bf A}_1 + \sigma \hat{\bf Y}\hat{\bf n}^T }{  V_g(\sigma +  \tilde{\Omega}_0) }  dA_{\bf Y} \right] \, {\bf v}.
\label{tras_eq}
\end{align}
Using the axial symmetry of the geometry, we recast Eq.~(\ref{tras_eq}) in cylindrical coordinates, with $\theta = \text{atan}2(y,x)$ and $z$ the coordinate along the axis of symmetry, as 
\begin{align}
\label{eq:cylindrical_characteristic_eq}
\eta \sigma {\bf v} = -\mathbb{D}_1 {\bf v} + \int \left( \frac{1}{\sigma + \tilde{\Omega}_0(z)}\mathbb{D}_2(z) + \frac{\sigma}{\sigma + \tilde{\Omega}_0(z)}\mathbb{D}_3(z)\right) dz \, {\bf v},
\end{align}
where we reinterpret $\tilde{\Omega}_0$ as a function of $z$. The matrices $\mathbb{D}_{1,2,3}$ are diagonal with $\mathbb{D}_{j}^{1,1} =\mathbb{D}_{j}^{2,2}$ with the superscript denoting the entries of the matrix. Because of this algebraic structure, we know that $\hat{x},\hat{y}, \hat{z}$ are the eigenvectors of the problem, where the first two will share the same eigenvalues. This allows us to reduce the nonlinear eigenvalue problem to the following two nonlinear algebraic equations:
\begin{align}
\eta \sigma_{i}  &= -\mathbb{D}^{i,i}_1  + \int \left(\frac{\mathbb{D}^{i,i}_2(z)}{\sigma_{i} + \tilde{\Omega}_0(z)} + \sigma_{i} \frac{ \mathbb{D}^{i,i}_3(z)}{\sigma_{i} + \tilde{\Omega}_0(z)}\right) dz, \quad i = 1,3.
\label{tras_eq_simple}
\end{align}
We numerically solve Eqs.~(\ref{tras_eq_simple}) and find that there are only two solutions per index.
%Through numerical experimentation, we observe that for each index in Eq. (\ref{tras_eq_simple}), there are only two solutions, which depending on the parameters chosen, can be in any quadrant of the complex plane. 
The first two eigenvalues are associated with four eigenmodes in the $xy-$plane and the remaining two correspond to motion in the $z$ axis. Importantly, each direction of motion is completely independent of the others, just as in the case of the sphere. In \S~\ref{subsec:Methods_geometric_intuition} we provide a geometric argument on the structure of these solutions.

We further compute the eigenvalues of Eqs.~(\ref{tras_eq_simple}) 
%(\ref{eq:dX1dt_general},\ref{eq:dP1dt_general}) 
for a family of spheroidal shapes $\Gamma_\alpha$, defined as $(x/a)^2+(y/a)^2+(z/c)^2=1$, with $\alpha=c/a$ as the shape factor. The parameter $a$ is computed such that the total area of the cell remains invariant as we vary $\alpha$ (Fig.~\ref{Fig3}A). Details of the discretization and eigenvalue computation are provided in \S~\ref{subsec:Methods_arbitrary_geometry}. We construct a phase diagram as a function of CFG number $M$ and shape factor $\alpha$ by classifying each point based on the eigenvalues and the corresponding centrosome movements in each direction: stable centering, oscillatory, and unstable.

For $\alpha=1$, which is a sphere, we find three regimes that we discussed earlier: for low $M$, the centrosome is stable in the center (case 1). As $M$ increases, the centrosome exhibits an oscillatory behavior (case 4), and for large enough $M$, the center becomes unstable (case 9) (Fig.~\ref{Fig3}B). For oblate spheroids with $\alpha<1$, we observe a more diverse set of behaviors where the centrosome exhibits different dynamics in different directions. In case 2, the centrosome is stable in the $xy$-plane but oscillates in $z$; in case 6, the centrosome is unstable in $z$ direction and oscillates in the $xy$-plane, and in case 7, the centrosome is unstable in $z$ and stable in the $xy$-plane (Fig.~\ref{Fig3}B). For prolate spheroids with $\alpha>1$, we find case 3 where the centrosome is stable in $z$ but oscillates in the $xy$-plane, a small parameter region that the centrosome is stable in $z$ and unstable in the $xy$-plane (case 8), and a region that centrosome oscillates in $z$ and is unstable in the $xy$-plane (case 5) (Fig.~\ref{Fig3}B). The diverse dynamics we observe in spheroids highlight the effect of cell geometry on centrosome motion.

\subsection{Nonlinear structures of the Hopf bifurcation \label{subsec:NonlinearStructureHopf}} 
%In this section, we explore the nonlinear behavior of the system under a variety of conditions. This includes investigating aster dynamics for large values of $M$, analyzing the impact of MT front dynamics on centrosome motion, and examining geometries with discrete rotational symmetries. Initially, our numerical simulations reveal that Hopf bifurcations occur within a constrained parameter range, exhibiting both subcritical and supercritical behaviors dependent on these parameters. Subsequently, we demonstrate the emergence of periodic attractors in systems with a freely moving centrosome, positing that these attractors represent energetically preferable states compared to on-axis oscillatory motions. 

\begin{figure}[t]
\includegraphics[keepaspectratio=true]{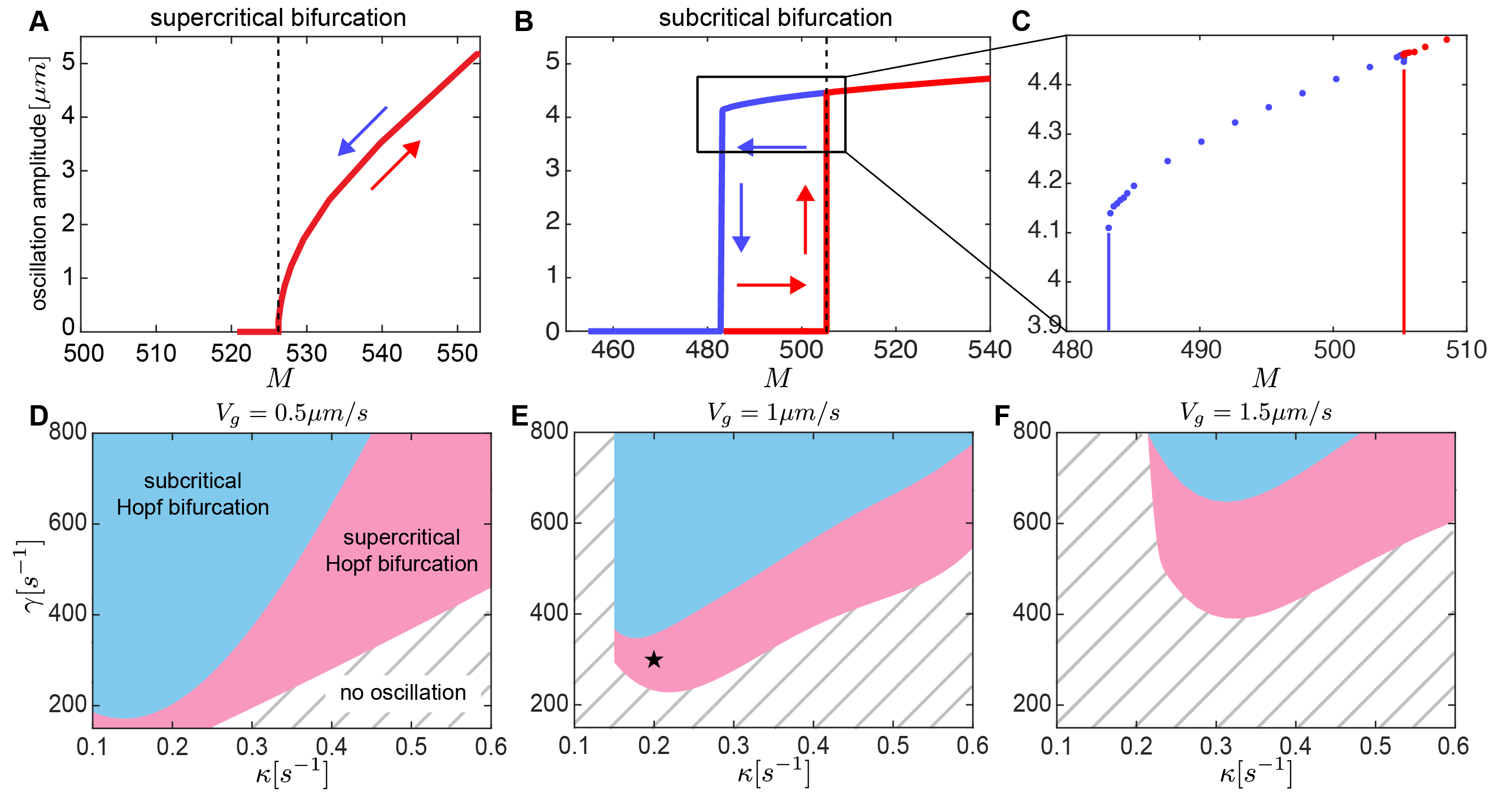}
\caption{Nonlinear dynamics of the centrosome near Hopf bifurcation.
(A) Oscillation amplitude as a function of $M$ for forward (red) and backward (blue) continuations using parameter values in Table~\ref{tab:parameters}. The dashed line indicates the value of $M_H$ at the Hopf bifurcation. (B) Oscillation amplitude as a function of $M$ for forward (red) and backward (blue) continuations using parameter values in Table~\ref{tab:parameters} with $\gamma=400 [s^{-1}]$. The dashed line indicates the value of $M$ at the Hopf bifurcation. (C) A zoom-in of panel B, where the dots represent the results of the simulation. (D-F) The type of the Hopf bifurcation is shown as a function of MT nucleation rate $\gamma$ and CFG detachment rate $\kappa$ using parameter values in Table~\ref{tab:parameters} for three values of $V_g=0.5, 1, 1.5 \mu m/s$. In the dashed region, the centrosome is stably centered. In the pink region, the centrosome exhibits supercritical Hopf bifurcation, and in the blue region, the Hopf bifurcation is subcritical. The star in E indicates parameter values of Table~\ref{tab:parameters}.}
\label{Fig5}
\end{figure}

Motivated by the form of the unstable eigenfunction for the spherical case, we study the nature of the Hopf bifurcation in the simplest case of an axisymmetric centrosome motion (along about the $z$-axis). To distinguish between sub- and supercritical Hopf bifurcations as a function of the bifurcation parameter $M$, we simulate the system by gradually varying $M$ around $M_H$ (the critical CFG number at the Hopf bifurcation). For a supercritical Hopf bifurcation, the transition from stationary equilibrium to oscillation is continuous in oscillation amplitude and history independent. This transition is discontinuous for a subcritical Hopf bifurcation, and the system exhibits hysteresis.

Taking advantage of the axial symmetry, we solve Eqs.~(\ref{eq:Zdot_3d}-\ref{eq:impingement_rate_3d}) using a time-dependent spectral code (see \S~\ref{subsec:spectral_method}). We expand the occupancy probability $P$ in Legendre polynomials and integrate the truncated system of nonlinear ODEs in time, allowing for fast and accurate simulations to sweep a large parameter space. Starting the simulation with CFG number $M < M_H$, we gradually increase $M$ past the Hopf bifurcation from the left (forward continuation). Once $M>M_H$, we continue this simulation by decreasing $M$ past the Hopf bifurcation from the right (backward continuation). A combination of forward and backward continuation simulations allows us to compute the hysteresis and, thus, determine the nature of the Hopf bifurcation. For example, for parameters in Table~\ref{tab:parameters} (star in Fig.~\ref{Fig5}E), there is no hysteresis, the Hopf bifurcation is supercritical, and the amplitude of oscillation continuously increases from zero as we pass through the Hopf bifurcation (Fig.~\ref{Fig5}A). Changing MT nucleation rate $\gamma$, from $300 s^{-1}$ to $400 s^{-1}$, while keeping other parameters unchanged, we find hysteresis in the model, the bifurcation becomes subcritical, and the oscillation amplitude abruptly changes passing through the bifurcation (Fig.~\ref{Fig5}B-C).

We perform a parameter sweep as a function of MT nucleation rate $\gamma$ and CFG detachment rate $\kappa$ to construct a phase diagram for the type of the Hopf bifurcation in the S-model (Fig.~\ref{Fig5}D-F). We find three regimes: For low values of detachment rate $\kappa$ and nucleation rate $\gamma$, the centrosome does not stably oscillate (Fig.~\ref{Fig5}D-F, stripped region). However, for sufficiently large $\kappa$ and $\gamma$, the centrosome oscillates. For a fixed $\kappa$, with increasing $\gamma$, the system transients to supercritical Hopf bifurcation (Fig.~\ref{Fig5}D-F, pink), followed by subcritical Hopf bifurcation (Fig.~\ref{Fig5}D-F, blue). While the structure of the phase diagrams preserves with varying $V_g$, the details of these transitions change.

\textbf{ \textit{ Comparison with experiment.}} Previous studies in \textit{C. elegans} revealed that depleting the gpr-1/2 genes, which impairs GPR-1/2 protein functionality, results in reduced cortical pulling forces and cessation of spindle oscillation~\cite{grill2003distribution}. A subsequent titration study~\cite{Pecreaux2006_CB} demonstrated a gradual decrease in spindle oscillation amplitude with the progressive depletion of gpr-1/2, identifying a critical force-generator number threshold below which spindle oscillation stops. Within the framework of the S-model, these findings align with the behavior expected from a supercritical Hopf bifurcation governing the \textit{C. elegans} spindle. Our model predicts oscillation amplitudes of approximately $\sim 5 \mu$m, similar to experimental observations ($\sim 5 \mu$m for maximum oscillation amplitude), and oscillation frequencies around $\sim 0.16 s^{-1}$ (similar to experiment, $\sim 0.25 s^{-1}$). Notably, the model confirms the occurrence of a supercritical Hopf bifurcation. Despite the model's simplification—comparing the \textit{C. elegans} spindle, represented as two connected centrosomes in an ellipsoidal geometry, to a single centrosome within a spherical cell—the theoretical and experimental results show remarkable concordance.

\section{Equatorial orbits and centrosome competition}
\label{EqOrbsThomson}

In this section, we explore the nonlinear behavior of the system under a variety of conditions. This includes investigating aster dynamics for large values of $M$, analyzing the impact of MT front dynamics on centrosome motion, and examining geometries with discrete rotational symmetries. We first demonstrate the emergence of periodic attractors in systems with a freely moving centrosome, positing that these
attractors represent energetically preferable states compared to on-axis oscillatory motions. Next, we present evidence that stoichiometry alone
can effectively position two centrosomes at diametrically opposite ends of the cell, mirroring the arrangement of spindle poles during mitosis. Finally, we demonstrate that multiple asters inside a sphere interact with each other through a competition-induced repulsion that can position themselves into symmetric configurations, as electrons constrained to a sphere and interacting via repulsive Coulomb potentials in the classical Thomson problem.
\subsection{Nonlinear orbital dynamics beyond $M_H$ \label{subsec:NonlinearOrbit}}
\begin{figure}[h!]
\includegraphics[scale = 0.93]{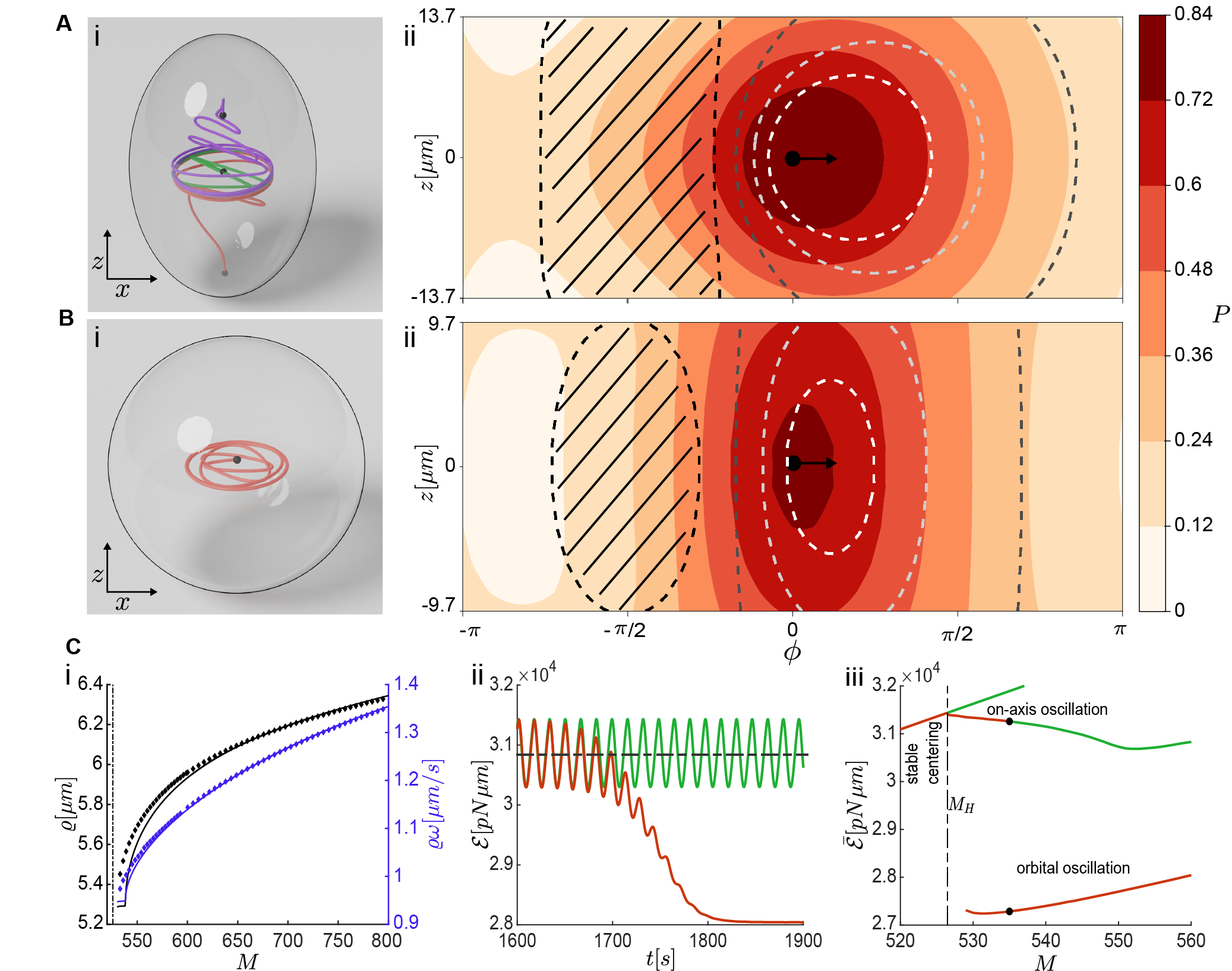}
\caption{Three-dimensional nonlinear centrosome dynamics in spheroids for parameters in Table 1. (A-B)i: Trajectories of a centrosome starting from different initial positions (black dots) in a sphere and prolate spheroid ($\alpha=1.5$) with $M=600$. (A-B)ii: The contour plots of steady-state surface occupancy probability $P$ (in color) and impingement rate $\Omega$ (dashed lines), as a function of azimuthal angle $\phi$ about the equator, and signed height along the $z$-axis. The patterned region shows where $\Omega=0~s^{-1}$, the black dot indicates the centrosome position, and the arrow shows the orbit direction. 
%Supplementary Videos 2 and 3 show the orbiting aster in a prolate spheroid and a sphere, respectively. 
(C)i: For case (B) of a sphere, the steady-state values of orbital radius $\varrho$ (black) and orbital speed $\varrho\omega$ (blue), as a function of $M$. Dots show the values from full simulations, while solid lines indicate solutions found numerically from a traveling wave Ansatz. The dashed line shows $M=M_H$. (C)ii: The energy ${\cal E}$ versus time for on-axis oscillation (green) and orbital oscillations (red). The horizontal dashed line indicates the time average of the energy for on-axis oscillation. (C)iii: The time-averaged energy as a function of $M$ (green, transient solutions; red, asymptotic steady states). Both on-axis and orbital solutions are stable for $M$ between $526$ (dashed line) and $535$ (dots).}
\label{Fig6}
\end{figure}

Here, we continue with the numerical analysis of the 3D nonlinear dynamics of the S-model aster, for which we will need to take account of the dynamics of the MT-front $\mathbf{S}$. We examine three distinct cell shapes: prolate spheroid ($\alpha=1.5$, similar to \textit{C. elegans} single cell embryo), sphere ($\alpha=1$), and oblate spheroid ($\alpha=0.5$), with all taken to have equal surface areas. All simulations are for $M=600>M_H$ ($M_H = 325, 415,525$, for $\alpha=0.5$, $1.5$, and $1$, respectively) and are prepared as follows. We initially fix the centrosome position at a random location $\mathbf{X}_0$ inside the cell with $\mathbf{S}$, the MT front position a sphere of radius zero at $\mathbf{X}_0$ (i.e., initially no MTs in bulk), and with $P(\cdot,t=0)\equiv 0$ (all CFGs unoccupied). Holding the centrosome in place, we evolve the occupancy probability $P$ via Eq.~(\ref{eq:dPdt_overlap}) until $P$ reaches an equilibrium $P_{\infty}$. In the language of Eq.~(\ref{eq:Zdot_3d}) at long times we are exerting an external force $\mathbf{f}^{ext}_\infty=-Mf_0\int_\Gamma P_\infty\hat{\boldsymbol{\xi}}dA$. We then release the centrosome to move while updating the MT front according to the algorithm described in \S~\ref{subsec:TrackingMT}. We performed 10 simulations for each cell shape, and show examples in Fig.~\ref{Fig6}.

For the prolate and spherical cases, we find that the centrosome can move into an internal equatorial orbit, which can be interpreted as a steady traveling wave solution of the S-model. Consider first the prolate case  (Fig.~\ref{Fig6}A). Our simulations show that, with near independence of initial position, the centrosome is attracted to the $xy$-plane (panel Ai) and thence to a circular orbit in that plane (initial data on the $z$-axis is attracted to the origin). Different initial data yield different directions of motion upon this attractor. The colorfield in panel Aii shows the translating $P$ field for steady motion on the orbital attractor (in the plot, moving rightwards). The dashed curves show contours of the impingement rate $\Omega$ whose motion leads the $P$ field. Their azimuthal asymmetry reflects the mechanism that drives this traveling wave. If more MTs are impinging on the right than on the left, as they are in Fig.~\ref{Fig6}Aii \& Bii, this creates a force imbalance on the centrosome that pulls it to the right, reinforcing the effect. Note that in this case the traveling wave speed along the cell surface is $\sim 1.3\; \mu m/s$, greater than the MT growth speed $V_g=1\; \mu m/s$ which means that in the aft of this wave, growing centrosomal MTs never reach the cortex (i.e. $\Omega$ is zero there; patterned region in Fig.~\ref{Fig6}Aii). 

For the case of the sphere (Fig.~\ref{Fig6}B) we again find attraction onto internal equatorial orbits, but since the sphere geometry make no selection of a particular equator, these attracting orbits depend upon initial conditions (Fig.~\ref{Fig6}Bi). The traveling wave fields for $P$ and $\Omega$ are again asymmetric, closely resembling those found in the prolate case, again reflecting the asymmetry in centrosomal MT impingement that drives the wave. 
We have numerically constructed the equatorial orbit solution by assuming a traveling wave structure for Eqs.~(\ref{eq:Zdot_3d},\ref{eq:dPdt_overlap}) which becomes a nonlinear eigenvalue-eigenfunction problem for the angular wave-speed $\omega$, centrosome orbital radius $\varrho$, and the $P$ field, and is solved via an iterative gradient method, see \S~\ref{subsec:OrbitingCentrosome}. We find good agreement between the long-time simulations and the numerical solution for exact equatorial orbits, and Fig.~\ref{Fig6}Bii shows the $P$ and $\Omega$ fields, as in Fig.~\ref{Fig6}Aii.

The equatorial attractors emerge essentially with the Hopf bifurcation. Figure~\ref{Fig6}Ci shows that $\varrho$ (black curve) has a weak dependence on $M$, as does the orbital speed $\varrho\omega$ (red curve). Returning to $M=600$, we consider the energy ${\cal E}$ for centrosomes started initially near the fixed point at cell center (panel Cii). The red curve is typical; the energy shows an immediate oscillation as the centrosome is oscillating along a line through the origin. After several oscillations the centrosome moves outwards and relaxes into an equatorial orbit of lower, and constant, energy. For comparison, the green curve shows the dynamics for a centrosome started upon the $z$-axis. Symmetries restrict its motion to oscillations along the axis, with its energy oscillating about the energy level of the fixed point itself (dashed line). Figure~\ref{Fig6}Ciii examines the behavior of ${\cal E}$ across three allowed types of dynamics: centrosome at the center, on-axis oscillation (along the $z$-axis), and moving on an equatorial orbit. Note that there is a region of bistability of the second and third cases, and that the equatorial orbit shows the lowest energy overall.

Finally, in the oblate cell, we only observed on-axis oscillation along the $z$ axis. This observation is consistent with the linear stability analysis in Fig.~\ref{Fig4}B, where the only unstable mode for an oblate spheroid is along the $z$ axis, while for prolate the most unstable modes are oscillations in the $xy$-plane.

\clearpage

\subsection{Centrosome competition and positioning\label{subsec:MultiCentrosomes}}
\begin{figure}[h!]
\includegraphics[keepaspectratio=true]{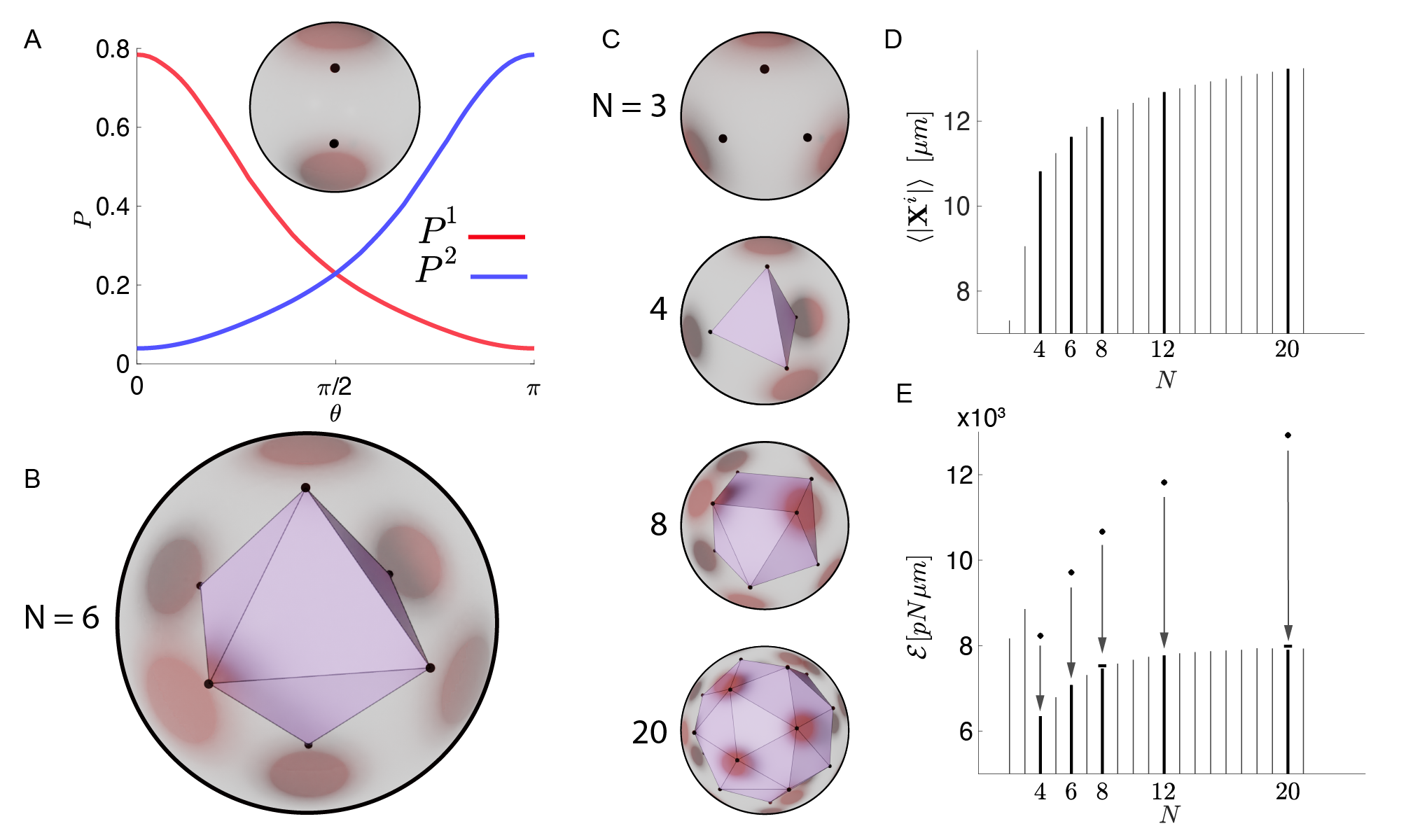}
\caption{Arrangement of multiple centrosomes inside a sphere for parameter values in Table 1 and $M = 100$. The surface color-field in (A-C) shows the total occupancy probability. (A) Steady state occupancy probabilities for two centrosomes in a spherical cell, as a function of the polar angle $\theta$ from the axis that aligns with the centrosomes. The inset shows the two centrosomes (black dots) inside of the cell. (B) Stable positioning of six centrosomes, forming an octahedron, inside the sphere.  Supplementary Video 2 shows the relaxation of six asters into the vertices of an octahedron. (C) Stable positioning of 3, 4, 8, and 20 centrosomes in the sphere.  (D-E) The average distance of centrosomes from the sphere center (cell radius is $W=15$) and total energy ${\cal E}$ for simulations in (A-C). Dots in E denote the total energy at the beginning of the simulation when all the centrosomes are placed near the center, bars over $N=8$ and $20$ represent the energy of the platonic solids, less than $1\%$ higher than the non-platonic equilibrium configurations. 
\label{Fig8}
}
\end{figure}

So far, we have discussed the positioning of single asters in cells. Prior to mitosis, however, the centrosome duplicates and the two newly formed asters position themselves at the two poles of the spindle, itself typically aligned along the long axis of the cell. The position and orientation of spindle poles then determine the position and orientation of the division furrow. In Farhadifar {\it et al.}~\cite{farhadifar2020stoichiometric}, we used a discrete formulation of the independent motor S-model for two asters and showed that it quantitatively explains the dynamics of spindle elongation and length determination in the embryos of \textit{C. elegans} and other nematodes. In a recent study, Fujii \textit{et al.} created a \textit{C. elegans} strain in which more than two centrosomal asters coexist within the same cell~\cite{fujii2024enucleation}. This strain is created by the deletion of the \textit{emb-27} and \textit{klp-18} genes, which yields an enucleated embryo with functional centrosomes. As the embryo progresses through its cycles, the centrosomes duplicate, but the mitotic furrow does not fully form, thus generating an embryo with multiple asters within the same cell. The authors observed that after each round of duplication, the new asters separate from each other and spread out throughout the cell. 

Inspired by these experiments, we performed simulations of multiple centrosomes ($N=2-21$) in a sphere for parameters in Table 1 and $M=100$. For these parameter values, a single centrosome is stably centered. In our S-model, note that centrosomes only interact through their stoichiometry-mediated competition for CFGs. That is, if a CFG is occupied by MTs of centrosome A it cannot be simultaneously occupied by MTs of centrosome B. This competition creates an effective repulsive interaction between asters: If two asters are nearby then the CFGs within their common reach will divide their pulling forces between the two asters, with that division determined by the relative impingement rates of each aster. This creates a force imbalance towards the directions of noncompetitive binding, that is, the centrosomes will move away from each other. A linear calculation for the stability of two centrosomes at the sphere center confirms this is an unstable configuration for any value of $M$, and two centrosomes always move away from the center in the opposite direction.

This is ably illustrated for the case of two centrosomes ($N=2$). To start, we hold the centrosomes fixed near each other close to the cell center, find the corresponding steady occupancy probability using Eq.~(\ref{eq:multicentrosomes_p}), and then release the asters. As Fig.~\ref{Fig8}A shows, the two centrosomes separate from each other and move to opposing sides of the sphere, with their pointwise CFG occupancy probabilities maximal near their corresponding pole and decreasing to small values further away. 
%{\bf MJS: It would be good here to have a comment on why the centrosomes do not go all the way to the wall.} 
Each centrosome does not go all the way to the cell periphery because when it is right by its pole, the pulling force toward its pole reduces to zero and the centrosome is pulled to move toward the opposite pole. If repeated within an ellipsoidal cell, the two asters will migrate towards opposite poles of the long axis, as demonstrated in \cite{farhadifar2020stoichiometric}. 
and seen in Fig.~\ref{Fig1_1}B.

Keeping the number of CFGs the same, we repeat this simulation with $N=6$ asters and find that the centrosomes spread outwards and move to the vertices of a octahedron, a platonic solid, with those vertices sitting on a sphere of radius $\rho_6=11.63 < W(=15$ $\mu$m); see Fig.~\ref{Fig8}B and Supplementary Video 2. The steady CFG occupancy probability for each centrosome is highest at the nearest cell surface point. This ultimate arrangement of asters is allowed by the spherical geometry and uniformity of the CFG distribution, as are many other possible final states (e.g. all asters sitting on the equator of an internal sphere whose radius is to be determined). That said, we find that generic initial data of centrosome positions (i.e. having no respected symmetry) relax to a  platonic octahedron on a sphere of radius $\rho_6$.

Interestingly, we find that the $N=4$, $6$, and $12$ cases also appear to be ``globally attracted" to their corresponding platonic solid; see Figs.~\ref{Fig8}C. Not so for the $N=8$ and $20$ cases where the platonic solids appears to have much more limited basins of attraction, and for which we also observe other apparent steady-states. For example, for $N=8$, if we start the simulation with asters on the vertices of a centered cube on a small sphere, it will remain a cube and expands radially until it reaches the sphere of $\rho_8$. More generally, we instead find relaxation to the square antiprism, a twisted cube, where two opposing square faces are rotated $45$ degrees relative to each other (see Supplementary Video 3, where we initiate dynamics from a very slightly twisted cube). For $N=20$, with general initial conditions, we found a non-platonic configuration that resembles the 20 vertex elongated square cupola. The higher the number of centrosomes, the closer they position to the cell periphery (Fig.~\ref{Fig8}D).  For other values of $N$, the asters are almost evenly positioned upon the surface of an internal sphere, with a variance of $10^{-3}$ in centrosome distance to the center. The mean centrosome distance to the center increases with $N$ (Fig.~\ref{Fig8}D), which corresponds to the plateau of total energy ${\cal E}$ at equilibrium (vertical bars in Fig.~\ref{Fig8}E), $20\%$ ($40\%$) lower than the initial energy for $N=4$ ($N=20$) centrosomes placed at the sphere center (diamonds in Fig.~\ref{Fig8}E).

We also investigated numerically the local stability of all the platonic solid cases ($N\in\{4,6,8,12,20\}$) by randomly perturbing the centrosome positions at the vertices of equilibrium platonic solids on a sphere of radius $\rho_N$. We first determine $\rho_N$ by simulating the relaxation of centrosomes, placed on the platonic solid verticies of radius $\rho^0_N<W$ at the beginning of the simulations, until they reach an equilibrium distance $\rho_N$ to the center. Throughout these simulations the centrosome configurations retain their platonic symmetries, and relax to an equilibrium distance $\rho_N$ with a variance less than $10^{-6}$. We then perturb the centrosome positions from the equilibrium platonic solid configurations. For $N=4$, $6$ and $12$, we observe the randomly perturbed configurations relax back to the platonic solid of radii $\rho_N$, suggesting that these are locally stable solutions. For $N=8$ and $N=20$, however, the perturbed configurations evolve to non-platonic shapes, suggesting that the cube ($N=8$) and dodecahedron ($N=20$) configurations are unstable solutions. Interestingly we also find that, for both $N=8$ and $N=20$, the platonic configuration has a total energy ${\cal E}$ (horizontal bars in Fig.~\ref{Fig8}E) that is slightly (less than $1\%$) higher than the non-platonic equilibrium configuration. 

%{\bf MJS: The discussion of effective repulsion goes here.} 
The ordering of centrosomes inside a sphere is indicative of an effective repulsion between centrosomes in the S-model. We can understand this heuristically in a simplified version of the S-model. As in Farhadifar {\it et al.} \cite{farhadifar2020stoichiometric}, assume that the binding probability is determined quasi-statically (i.e. $P^i_t=0$) and that ${\cal I}(P)=1-P$ (i.e., the independent motor model). In this case, the occupancy probabilities can be found exactly in terms of the impingement rates as $P^i(\mathbf{Y}) = \Omega^i(\mathbf{Y})/(\kappa+\bar\Omega(\mathbf{Y}))$ where $\bar\Omega = \Sigma_{i=1}^N \Omega^i$, that is, the total impingement rate from all centrosomes at a surface point $(\mathbf{Y})$. In this approximation we then have
\begin{equation}
\label{eq:Repulsion}
    \eta\dot{\mathbf{X}}^i =Mf_0 \int_\Gamma \frac{\Omega^i(\mathbf{Y})}{\kappa+\bar\Omega(\mathbf{Y})} \hat{\boldsymbol{\xi}}^i(\mathbf{Y}) \rho({\bf Y})dA_\mathbf{Y}~.
\end{equation} 
Thus, the direction of centrosome motion is set by a weighted integral average of pulling force vectors $f_0 \hat{\boldsymbol{\xi}}^i(\mathbf{Y})$, with regions of high population impingement rate divisively reducing that region's contribution to the total pulling force on the centrosome. Consequently, centrosomes will be pulled preferentially by their respective regions of relatively unshared CFGs. This is, in effect, a competition-induced repulsion between asters and is consistent with the linear stability calculation that shows two centrosomes at the sphere center always move away from each other.

The results found here very closely track those of the famous Thomson problem for energy-minimizing configurations of electrons constrained to a sphere and repelling each other through Coulombic forces \cite{tomson1904structure}. The minimizing equilibrium configurations of up to $470$ electrons are available in the literature \cite{atiyah2003polyhedra}. 
For $4$, $6$ and $12$ electrons on a sphere, the corresponding platonic solids' vertices (tetrahedron, octahedron, and icosahedron, respectively) are the equilibrium positions that minimize the total Coulombic potential. Eight electrons settle to the square antiprism, similar to the antiprism of eight asters in Fig.~\ref{Fig8}C but with a different aspect ratio.
Twenty electrons equilibrate to the twenty vertex non-regular polyhedron with $D_{3h}$ symmetry.
For both $N=8$ and $20$, the distribution of pairwise distances between repelling electrons is different from that between asters in Fig.~\ref{Fig8}C. 
For $N=4$, $6$, and $12$, the equilibrium relative orientations of electrons in the Thomson problem are those found here, though with the aster choosing their own sphere upon which to order. Thus, the spatial ordering of centrosomal asters inside a spherical cell, at least as modeled here, seems a new, generalized Thomson problem \cite{smale1998mathematical,bowick2002crystalline}, though in a nonconservative system where particles interact indirectly through competition, and move through a balance of active forces and dissipation. Finally, note that in our simulations, CFGs are uniformly distributed across the cell surface, and presumably then, due to this symmetry, the asters (nearly) evenly distribute.

% Many cells have more than two centrosomes -- e.g., {\it C. elegans} embryos have two centrosomes that constitute the spindle body poles. Previously, we showed that two centrosomes move to opposite sides of the cell using a reduced S-model for which the temporal dynamic of the binding probability $P$ was ignored \cite{farhadifar2020stoichiometric}.  Another example of multiple centrosomes is in the early development of \textit{Drosophila melanogaster} embryos, where multiple nuclei and centrosomes inside a single cell reach the cortex with remarkable positional order. In \textit{giant nuclei (gnu)} mutant embryos, where there are many more free centrosomes (asters) than nuclei, the centrosomes self-organize to form lattice structures as nuclei do in the wild-type \textit{Drosophila} embryos \cite{deCarvalho2022asterrepulsion}. In explants of a \textit{gnu} mutant embryo, these centrosomes are confined in a circular droplet and form lattices with a distribution of inter-centrosome distances that depends on the total number of centrosomes. A short-range repulsion between centrosomes is hypothesized to explain the observed spatial order of centrosomes \cite{deCarvalho2022asterrepulsion}. 

\section{Discussion\label{sec:discussion}}

Cortical pulling forces play a fundamental role in the positioning and orientation of the mitotic spindle, perhaps best illustrated during asymmetric division, where the spindle accurately positions off the cell center~\cite{knoblich2008mechanisms, gonczy2008mechanisms, morin2011mitotic}. It has been argued that cortical pulling forces are destabilizing, and these forces alone are not sufficient to position the spindle in cells~\cite{howard2006elastic}. Indeed, a set of models have been developed to explain spindle positioning using mechanisms such as MT polymerization-driven pushing forces against the cell cortex, pulling forces from motor proteins carrying payloads along MTs, MT length-dependent pulling forces from cortical motors, and forces from microtubule friction with the cell wall~\cite{tran2001mechanism, grill2003distribution, grill2005theory, howard2006elastic, CampasSens2006_PRL, thery2007experimental, kozlowski2007cortical, hara2009cell, zhu2010finding, shinar2011model, laan2012cortical, pavin2012positioning, ma2014general, garzon2016force, pecreaux2016mitotic, letort2016centrosome, howard2017physical, tanimoto2018physical}. Alternatively, a simple, intuitive idea was suggested that stable centering of the aster could be accomplished by pulling forces if there are fewer CFGs than astral MTs \cite{Grill2005_DC}. The stoichiometric model naturally expresses this possibility through its bottom-up development and, in its earlier forms, quantitatively explains spindle positioning, elongation dynamics, and scaling with cell size among nematodes~\cite{farhadifar2020stoichiometric}, and quantitatively explains the oscillatory dynamics of the spindle in \textit{C. elegans}~\cite{WuEtAl2023}. A version of the independent motor model, but with a force-dependent detachment rate, was previously used to model chromosome oscillation in human cells~\cite{CampasSens2006_PRL}.

% Stoichiometric effects are also essential to the dynamics of mitotic choromosome, giving rise to a sustained periodic oscillation from the competition between the kinetochore and chromokinesin motors on the chromosome arms \cite{CampasSens2006_PRL}.

Here we developed an elaborated S-model that accounts for CFG binding domain overlap, which explains recent experiments, and a generalized dynamical formulation. We used this model to explore the mathematical and dynamical structure of how centrosomal asters interact with CFG populations, focusing particularly on the effects of motor density, cell shape, and centrosome number. The system is surprisingly rich in behaviors. As we describe below, this study creates the basis for the modeling of yet more complex, and fundamental, intracellular dynamics during development. 

Previous \textit{in vitro} and \textit{in vivo} studies have demonstrated that, in some cases, aster positioning is governed by a dynamic equilibrium between the pulling forces exerted by motor proteins and the pushing forces arising from microtubule polymerization against the cell cortex~\cite{meaders2020pushing, tanimoto2016shape-motion, laan2012cortical, sulerud2020microtubule}. As microtubules elongate and encounter the cell periphery, they push on the centrosomal aster, potentially leading to microtubule buckling~\cite{dogterom1997measurement}. Computational tools such as Cytosim~\cite{nedelec2007collective} and SkellySim~\cite{nazockdast2017fast,dutta2024self} have been employed to simulate these phenomena, accommodating extensive microtubule deformations and in the case of SkellySim, incorporating hydrodynamic interactions due to microtubule motion. Various coarse-grained models have also been constructed to account for microtubule pushing forces, assuming microtubules grow briefly against the cell surface and push, but disassemble before buckling ~\cite{zhu2010finding, howard2006elastic, pavin2012positioning, MaLaanDogterom2014_NJP}. In line with such assumptions, pushing forces can be easily accommodated within the S-model, with the local pushing force being proportional to the microtubule impingement rate $\Omega$, which is part of the S-model. Out of curiosity and motivated by interesting {\it in vitro} experiments~\cite{laan2012cortical}, we studied the positioning, via the S-model, of an aster in a flat cuboidal geometry with motors populating only the sides. While it is clear that pushing must play a role there -- experimentally the aster centers in the lack of motors -- we were able to reproduce substantial parts of the results solely by pulling forces (see Supplementary Fig.\ref{Fig9}). 

The S-model is applicable to other biological phenomena involving aster positioning. For example, the life of \textit{C. elegans}, and many other metazoans, begins with two paternally delivered centrosomal asters, which will eventually fuse the male and female pronuclei, positioned at opposite ends of the embryo. Thus begins an orchestrated movement of asters and pronuclei starting with the centrosomes' migration toward opposing poles of the male pronucleus. Following this, the female pronucleus advances toward the male, leading to their fusion, and the resultant complex of asters and pronuclei collectively relocates to the cell center and reorients to align with the cell's axis. Similarly, in later stages, newly duplicated centrosomes and their asters migrate to opposite sides of the nucleus and orient themselves in the direction of the future spindle. Studies have shown that SUN-KASH protein complexes sitting within the nuclear envelope can bind motor proteins such as dynein and through this will exert a pulling force on centrosomal microtubules and the aster itself. Unlike the cell cortex, which is more fixed, the nucleus is mobile and can reposition and deform due to these pulling forces. It is straightforward to expand the S-model to account for centrosome/nucleus positioning and we are doing that now. There are new issues to confront such as overall force balance, and the ``shadows" cast by the nucleus that capture microtubules that might have otherwise impinged upon the cortex. Other extensions to the S-model including elastic responses such as from protein tether stretching (which can induce torques), a force/growth-velocity relationship, and hydrodynamic interactions \cite{SS2019}. 

The correspondence of multiple centrosome positioning, which is a damped and driven competition for motors, and the conservative and classical Thomson problem is amusing and surprising. We note that there have been a number of examples of damped and driven active matter systems behaving essentially as conservative systems; see, e.g. \cite{OSS2019,SO2023}. While a nonuniform distribution of CFGs is important to spindle placement, we did not study it here. Such nonuniformity has been posited to explain clustering of centrosomes~\cite{mercadante2023cortical}, and it would be interesting to understand what the the S-model predicts for aster arrangements in cells with non-uniform CFG distributions. 

Finally, in this paper, we also examined the role of cell shape. Understanding how cell geometry regulates spindle positioning and orientation is fundamental for understanding embryonic development. For example, in \textit{C. elegans}, embryogenesis progresses from a single cell encapsulated in a relatively rigid shell to about a thousand cells by successive cell divisions, leading to changes in cell volume and shape. During each division, the spindle forms and precisely orients along the long axis of the cell, often referred to as Hertwig's rule in developmental biology~\cite{hertwig1884einfluss}. The biophysical mechanism behind Hertwig's rule has been actively pursued~\cite{pierre2016generic, bosveld2016epithelial, middelkoop2023cytokinetic}, and we are now testing predictions of the S-model for spindle positioning and its 3D orientation during \textit{C. elegans} development, where we observed quantitative agreement with experiments. 

\section*{Acknowledgments} The authors acknowledge fruitful discussions with Daniel Needleman and Jane Wang. YNY acknowledges support by the National Science Foundation (NSF) under award DMS-1951600, and of the Flatiron Institute, part of Simons Foundation. MJS acknowledges support by NSF under award DMR-2004469. Numerical computations for this work were performed at facilities supported by the Scientific Computing Core at the Flatiron Institute. The microscopy image in Fig. 1, and associated Supplementary Video 1, were obtained at the CCBScope Observatory which is part
of the CCB$_{\mbox X}$ program.

\section*{Author Contributions}
All authors conceptualized the work; YNY, VGH and MJS conducted the analysis; YNY and VGH conducted numerical simulations; VGH designed the patch-based quadrature for spheroids; VGH and HZH derived the overlap motor model; VGH, RF and YNY made the graphics; MJS supervised the research; all authors contributed to the writing.  
\bibliography{CGOneCentrosome}

% %%
\begin{table}[h]
\begin{center}
    Biophysical parameters of stoichiometric model
\end{center}
%\caption*{Biophysical parameters of Stoichiometric Mode}
\begin{center}
\begin{tabular}{@{}lcll@{}}
       \toprule
       Biophysical Parameters & Symbol & Value & References \\
       \hline
%       \cmidrule(r){1-1}\cmidrule(lr){2-2}\cmidrule(l){3-3}
%        Microtubule nucleation rate &  $\dot n$ & $10^3/3 s^{-1}$\\
       spherical cell radius & $W$ & $15$ $[\mu \text{m}]$ & \cite{WuEtAl2023} \\
       microtubule growth rate      &   $V_g$   & $1\;[\mu \text{m}\;s^{-1}]$ $(0.5\sim 1.5$  $\mu\text{m}\;s^{-1})$ & \cite{srayko2005identification}\\
       microtubule catastrophe rate & $\lambda$ & $0.05\;[s^{-1}]$ $(0.025\sim 0.05\;s^{-1})$ &\cite{kozlowski2007cortical}\\
       microtubule nucleation rate & $\gamma$ & $300\;[s^{-1}]$ $(300\sim500\; s^{-1})$ & \cite{farhadifar2020stoichiometric}\\
%       Total number of microtubules & $N_{\infty}$ & $6000$ $(5\times 10^3\sim10^4)$ \\
%       Microtubule average length & $l_c$ & $20$ $\mu \text{m}$ $(15\sim 20$ $\mu \text{m})$ \\
       force-generator detachment rate & $ \kappa$  & $0.2\;[s^{-1}]$ $(0.1\sim 0.5\;s^{-1})$ &\cite{Redemann2010_PLoS} \\
       force-generator radius &  $r_m$ &  $1\;[\mu\text{m}]$ $(1\sim 1.5$ $\mu\text{m})$ & \cite{farhadifar2020stoichiometric}\\
       force-generator pulling force   &  $f_0$  & $10\; [p\text{N}]$ $(5\sim 10\;p$N) & \cite{kozlowski2007cortical}\\
       centrosome drag       &  $\eta$  &  $150\;[p\text{N} \;s\;\mu\text{m}^{-1}]$ $(150\sim 200\; p\text{N}\;s\;\mu\text{m}^{-1}$) &\cite{garzon2016force}\\
       \bottomrule
\end{tabular}
\end{center}
       \caption{Characteristic microtubule length $l_c=V_g/\lambda$, and number of microtubules at equilibrium $N_{\infty} = \gamma/\lambda = \gamma l_c/V_g$.\label{tab:parameters}}
\end{table}
%\clearpage
%
\section{Supplemental Material}

\subsection{Supplementary Videos}
\label{subsec:supp_videos}

\begin{itemize}
%[label=(\arabic*):]
\item Supplementary Video 1: First mitotic spindle in {\it C. elegans}
embryo. 
%\item Supplementary Video 2: Equatorial orbiting of a centrosomal aster in a prolate spheroidal cell with $\alpha=1.5$. 
%\item Supplementary Video 3: Equatorial orbiting of a centrosomal aster in a spherical cell.
\item Supplementary Video 2: Relaxation of six asters into the vertices of an octahedron inside a sphere.

\item Supplementary Video 3: Relaxation of eight asters into the vertices of a square antiprism inside a sphere. Initially placed at the vertices of a slightly rotated small cube, the centrosomes move radially first, and then rotate and relax to the vertices of a square antiprism.

\end{itemize}

\subsection{Algorithm for updating the MT front}
\label{subsec:TrackingMT} 
In \S~\ref{sec:CG1} we provided a general formulation of three-dimensional MT front dynamics inside a cell surface $\Gamma$. Here we provide details of evolving the front of leading MT plus-ends in terms of a coordinate system centered on the centrosome with polar angle $\varphi'(t)\in[0,\pi]$ and azimuthal angle $\theta'(t)\in[0,2 \pi)$. We focus on a convex cell surface $\Gamma$ so there is a unique pair of coordinates $(\varphi'(t),\theta'(t))$ that corresponds to coordinates of the force generators ${\bf Y}$ on $\Gamma$. We assume that the centrosomal array does not rotate, and represent the location of the MT plus-end (MT front) as ${\bf S}={\bf S}(\varphi'(t),\theta'(t)) = {\bf X}(t) + D_{\bf S}(\varphi'(t),\theta'(t))\hat{\boldsymbol{\xi}}(\varphi'(t),\theta'(t))$, where $D_{\bf S}$ is the distance from the centrosome to the MT front along the direction $\hat{\boldsymbol{\xi}}$, pointing from the centrosome to CFGs at ${\bf Y}(\varphi',\theta')$ at a distance $D = |{\bf Y}-{\bf X}|$. The relation between $D_{\bf S}$ and $D$ determines the state of the MT front; either

\noindent
{\bf(i)} the front reaches the cortex, i.e. $D_{\bf S}(\varphi'(t),\theta'(t))\ge D(\varphi'(t),\theta'(t))$. Centrosomal MTs are impinging on force generators in this direction, which implies $\Omega(\varphi'(t),\theta'(t))>0$; or

\noindent
{\bf(ii)} the front grows toward the cortex, i.e. $D_{\bf S}(\varphi'(t),\theta'(t))<D(\varphi'(t),\theta'(t))$ and $\partial_t D_{\bf S} = V_g$. Centrosomal MTs are not impinging on force generators, which implies $\Omega(\varphi'(t),\theta'(t))=0$.

In the following we summarize the numerical algorithm for evolving the three-dimensional centrosome motion with a MT front.  The superscripts denote the step in the time marching, starting from initial data of the occupancy probability $P(t^0)$, centrosome position ${\bf X}(t^0)$, and the unit vector $\hat{\boldsymbol{\xi}}(t^0)$ (from the initial centrosome position to a surface point ${\bf Y}$)  and the MT front position ${\bf S}(\varphi'(t^0),\theta'(t^0))$.

\begin{enumerate}
%[label=(\arabic*):]
\item Given a time-step $\triangle t$, and $({\bf X}(t^n)$, $P(t^n))$, at time $t^n = n \triangle t$ for $n\ge 0$, compute ${\bf X}(t^{n+1})$ as described in \S~\ref{subsec:simulation_methods}, and update the centrosome coordinates $(\varphi'(t^{n+1}),\theta'(t^{n+1}))$ and the distance $D(\varphi'(t^{n+1}),\theta'(t^{n+1}))$.

%\hat{\boldsymbol{\xi}}

\item Compute $\tilde D_{\bf S}(\varphi'(t^n),\theta'(t^n)) = D_{\bf S}^n(\varphi'(t^n),\theta'(t^n))+\Delta t V_g$; then compute $\tilde D_{\bf S}(\varphi'(t^{n+1}),\theta'(t^{n+1}))$ by interpolating $\tilde D_{\bf S}(\varphi'(t^n),\theta'(t^n))$ at the new coordinates (target points) $(\varphi'(t^{n+1}),\theta'(t^{n+1}))$. 

\item Update $D_{\bf S}$ and the impingement rate $R$ according to the following scheme: 
    \begin{itemize}
        \item 

        \noindent{If $\tilde D_{\bf S}(\varphi'(t^{n+1}),\theta'(t^{n+1})) \ge D(\varphi'(t^{n+1}),\theta'(t^{n+1}))$}
        
        \noindent{\hspace{0.4cm}$D_{\bf S}(\varphi'(t^{n+1}),\theta'(t^{n+1})) =D(\varphi'(t^{n+1}),\theta'(t^{n+1}))$}

        \noindent{\hspace{0.4cm}$\Omega(\varphi'(t^{n+1}),\theta'(t^{n+1})) = \left[{\bf V}_S\cdot\hat{\bf n} \right]_+ \chi\left(\frac{r_m}{D}\right)\left(\frac{\gamma}{V_g} e^{-D/l_c}\right)$}

        \item 

        \noindent{Otherwise ($\tilde D_{\bf S}(\varphi'(t^{n+1}),\theta'(t^{n+1})) < D(\varphi'(t^{n+1}),\theta'(t^{n+1}))$)}

        \noindent{\hspace{0.4cm}$D_{\bf S}(\varphi'(t^{n+1}),\theta'(t^{n+1})) = \tilde D_{\bf S}(\varphi'(t^{n+1}),\theta'(t^{n+1}))$}

        \noindent{\hspace{0.4cm}$\Omega(\varphi'(t^{n+1}),\theta'(t^{n+1}))=0$}  
    \end{itemize}
\end{enumerate}
We then update the MT front position to ${\bf S}(\varphi'(t^{n+1}),\theta'(t^{n+1}))$, and use the updated impingement rate $\Omega(\varphi'(t^{n+1}),\theta'(t^{n+1}))$ to compute $P^{n+1}$ as described in \S~\ref{subsec:simulation_methods}:
    \begin{align}
        \frac{P(t^{n+1})-P(t^{n})}{\Delta t} &= \Omega(\varphi'(t^{n+1}),\theta'(t^{n+1})) {\cal I}(P(t^{n})) - \kappa P(t^{n+1}),\nonumber\\
        \label{eq:Pnp1}
        P(t^{n+1}) &= P(t^{n}) + \frac{\Delta t}{1+\kappa\Delta t}\left(\Omega(\varphi'(t^{n+1}),\theta'(t^{n+1})) {\cal I}(P(t^{n})) - \kappa P(t^{n})\right).
    \end{align}
    This completes one time-step, and the iteration repeats until the end of the simulation.

\subsection{Spectral formulation for the axi-symmetric centrosome dynamics in sphere\label{subsec:spectral_method}}
Here we recast Eqs~(\ref{eq:Zdot_3d}-\ref{eq:impingement_rate_3d})  with axisymmetry ($z$-axis as the axis of symmetry) in Legendre polynomials.
  The centrosome moves along the $z$-axis with a position $z(t)$, and the occupancy probability is a function of the polar angle $\varphi\in[0,\pi]$.  We set the areal density $\rho\left({\bf Y}\right) = 1/|\Gamma|$.
The equations of motion for the axisymmetric system are
\begin{align}
\label{eq:Zdot0}
\eta \frac{dz}{dt} &= \frac{M f_0}{2}\int^{\pi}_0 d\varphi \sin\varphi \frac{W\cos\varphi-z}{D(\varphi,z)}P(\varphi,t),\\
\label{eq:Pdot0}
\frac{\partial P(\varphi,t)}{\partial t} &=  \Omega(\varphi,z)\left[\frac{1-e^{-\frac{M}{M_0}(1-P(\varphi,t))}}{\frac{M}{M_0}}\right]- \kappa P(\varphi,t),\\
%\frac{\partial P}{\partial t} &=  R(\phi,t)\left(1-P(\phi,t)\right)-kP(\phi,t),\\
%\label{eq:D0}
%D(\phi,z) &= \sqrt{1-2 z\cos\phi + z^2},\\
\label{eq:R0}
\Omega(\varphi,z) &= \left[\dot z \cos\varphi + \frac{V_g\left(W-z\cos\varphi\right)}{D(\varphi,z)}\right]_+e^{-\frac{D(\varphi,z)}{l_c}}\chi\left(\frac{r_m}{D(\varphi,z)}\right)\frac{\gamma}{V_g},
%\label{eq:chi0}
%\chi(\rho) &=\frac{1}{2}\left(1-\frac{1}{\sqrt{1+\rho^2}}\right),
\end{align}
where $D(\varphi,z) = \sqrt{W^2-2 W z\cos\varphi + z^2}$.
We expand the occupancy probability $P(\varphi,t)$ in Legendre polynomials $P_n(x)$ as $P(\varphi,t) = \sum^{\infty}_{i=0} c_i(t) P_i(\cos\varphi)$, where $P_i(x=\cos\varphi)$ is the $i$th order Legendre polynomial. To recast
equations~(\ref{eq:Zdot0})-(\ref{eq:Pdot0}) into a system of nonlinear equations in $z(t)$ and the coefficients $c_i(t)$, we first
expand the impingement rate in Legendre polynomials as
\[ \Omega(x,z)=\left(\frac{dz}{dt} f_1(x, z) + f_2(x, z)\right)\frac{\gamma}{V_g} = \sum^{\infty}_{i=0} \left(\frac{dz}{dt} f_{1i} + f_{2i}\right)  P_i(x)\frac{\gamma}{V_g},\]
where $P_i(x)$ is the $i$th order Legendre polynomial, and
\begin{align}
\label{eq:f1}
f_1 &\equiv x e^{-\frac{D}{l_c}}\chi\left(\frac{r_m}{D}\right) \equiv \sum^{\infty}_{i=0}f_{1i} P_i(x),\\
\label{eq:f2}
f_2 &\equiv \frac{V_g(W- z x)}{D}e^{-\frac{D}{l_c}}\chi\left(\frac{r_m}{D}\right) \equiv \sum^{\infty}_{i=0}f_{2i} P_i(x).
\end{align}
%
% As in the linear analysis we expand the occupancy probability $P$
% %~\ref{eq:dPdt_Rexpanded}
% in Legendre polynomials as $P(x,t)= \sum^{\infty}_{i=0} c_i(t) P_i(x)$. 
Here we  truncate at $c_1$ to obtain
a system of nonlinear equations for $z(t)$, $c_0(t)$ and $c_1(t)$:
\begin{align}
\label{eq:dZdt_fullR}
\eta\frac{dz}{dt} &= \frac{M f_0}{2}\left[-\frac{4}{3} c_0 \frac{z}{W} + \left(\frac{2}{3} - \frac{2}{5}\left(\frac{z}{W}\right)^2\right)c_1\right],\\
\label{eq:dc0dt_fullR}
\frac{dc_0}{dt} &= \Omega_0 - \kappa c_0 -\left(c_0 R_0 + \frac{1}{3} c_1 R_1\right),\\
\label{eq:dc1dt_fullR}
\frac{dc_1}{dt} &= \Omega_1 - \kappa c_1 - \left(c_0 R_1 + c_1 R_0 + \frac{2}{5} c_1 R_2\right),
\end{align}
with $\Omega_i \equiv \left(\frac{dz}{dt} f_{1i} + f_{2i}\right)\frac{\gamma}{V_g}$.
For $i=0$ and $i\ge 2$, the linearized equation of $c_i$ simply yields decaying dynamics:
\begin{align}
    \frac{d c_i}{dt} &= - \left(\kappa + I_0\left({\bar P}\right)\right) c_i,\\
    I_0\left({\bar P}\right) &= e^{-W/l_c}\chi\left(\frac{r_m}{W}\right)\gamma \frac{1-e^{-\frac{M}{M_0}(1-{\bar P})}}{\frac{M}{M_0}} >0,
\end{align}
where ${\bar P}$, the occupancy probability of the base state, satisfies the equation ${\bar P} = \frac{I_0\left({\bar P}\right)}{I_0\left({\bar P}\right) + \kappa}$.
 These results show that the linear instability of a centrosome in a sphere in \S~\ref{subsec:stability_in_sphere} involves only $c_1$ and $z$, and all the other modes are linearly stable and do not contribute to the instability.

\subsection{Centrosome orbiting around sphere center\label{subsec:OrbitingCentrosome}}
Here we focus on  a centrosome orbiting at a constant speed and a constant distance to the sphere center. In spherical coordinates, with the origin at the sphere center, the centrosome position 
${\bf X}_c = \varrho(t)(\sin\theta_c(t)\cos\phi_c(t)\hat{\bf e}_1 + \sin\theta_c(t)\sin\phi_c(t) \hat{\bf e}_2 + \cos\theta_c(t)\hat{\bf e}_3)$ and the surface of the sphere of radius $W$ has coordinates
${\bf Y} = W(\sin\theta\cos\phi\hat{\bf e}_1+\sin\theta\sin\phi\hat{\bf e}_2+\cos\theta\hat{\bf e}_3)$, where the polar angles 
$\theta_c, \theta \in [0, \pi]$, and the azimuthal angles $\phi_c, \phi \in [-\pi,\pi]$.
The governing equations for the centrosome dynamics in spherical coordinates (with areal density $\rho\left({\bf Y}\right) = 1/|\Gamma|$) are
\begin{align}
\eta\dot\varrho &= \frac{M f_0}{|\Gamma|}\int\frac{P(\theta,\phi,t)}{D}\left(-\varrho + W \cos\theta\cos\theta_c + W\sin\theta\sin\theta_c\cos(\phi-\phi_c)\right) W^2 \sin\theta d\theta d\phi,\\
\eta\varrho\dot\theta_c &= \frac{M f_0}{A}\int\frac{P(\theta,\phi,t)}{D} \left(\cos\theta_c\sin\theta\cos(\phi-\phi_c) - \cos\theta\sin\theta_c\right) W^3\sin\theta d\theta d\phi,\\
\eta\varrho\sin\theta_c \dot\phi_c &= \frac{M f_0}{A}\int\frac{P(\theta,\phi,t)}{D} \sin\theta\sin(\phi-\phi_c)W^3 \sin\theta d\theta d\phi,\\
\frac{\partial P}{\partial t} &= \left[ \dot\varrho\left(\cos\theta\cos\theta_c + \sin\theta\sin\theta_c\cos(\phi-\phi_c)\right) + \right. \\
& \left. \varrho\dot\theta_c (\cos\theta_c\sin\theta\cos(\phi-\phi_c)-\sin\theta_c\cos\theta)+ \varrho\sin\theta_c\dot\phi_c\sin\theta\sin(\phi-\phi_c)+\right. \nonumber \\
&\left. \frac{V_g}{D}(W-\varrho\cos\theta\cos\theta_c - \varrho\sin\theta\sin\theta_c\cos(\phi-\phi_c))\right]_+ \frac{\chi e^{-D/l_c}}{V_g}\gamma I(P) - \kappa P,  \nonumber  \\
D &=\sqrt{W^2+\varrho^2-2W\varrho\cos\theta\cos\theta_c-2 W \varrho\sin\theta\sin\theta_c\cos(\phi-\phi_c)},\\
\chi &= \frac{1}{2}\left(1-\frac{1}{\sqrt{1+\left(\frac{r_m}{D}\right)^2}}\right),\;\;\;
I(P) = \frac{1-e^{-\frac{M}{M_0}(1-P)}}{M/M_0}.
\end{align}
We focus on the orbiting motion of a centrosome in the $xy$-plane (where $\theta_c = \pi/2$) with a constant radius $\varrho$ from the sphere center and a constant angular frequency $\dot\phi_c = \omega$. With the change of variable $\bar\phi = \phi-\phi_c = \phi-\omega t$ and the assumption that $P(\theta,\phi,t)=P(\theta,\bar\phi)$, we obtain the following equations

\begin{align}
\label{eq:drhodt=0}
0 &= \frac{M f_0}{A}\int\frac{P(\theta,\bar\phi)}{D}\left(-\varrho  + W\sin\theta\cos(\bar\phi)\right) W^2 \sin\theta d\theta d\bar\phi,\rightarrow \varrho  = W\frac{I_2}{I_1},\\
\label{eq:omega0}
\eta\varrho\omega &=  \frac{M f_0}{A}\int\frac{P(\theta,\bar\phi)}{D} \sin\theta\sin(\bar\phi)W^3 \sin\theta d\theta d\bar\phi,\rightarrow \eta \varrho \omega =\frac{M f_0}{A}I_3,\\
\label{eq:dPdphi}
-\omega\frac{\partial P\left(\theta,\bar\phi\right)}{\partial\bar\phi}&= \left[\varrho\omega\sin\theta\sin(\bar\phi) + \frac{V_g}{D}(W -\varrho\sin\theta\cos(\bar\phi))\right]_+ \frac{\chi e^{-D/l_c}}{V_g}\gamma I(P) - \kappa P,\\
D &=\sqrt{W^2+\varrho^2-2 W \varrho\sin\theta\cos(\bar\phi)},
\end{align}
where 
\begin{equation}
\label{eq:I1}
I_1 = \int \frac{P(\theta,\bar\phi)}{D}\sin\theta d\theta d\bar\phi,
\end{equation}
\begin{equation}
\label{eq:I2}
I_2 = \int \frac{P(\theta,\bar\phi)}{D} \sin^2\theta\cos(\bar\phi)d\theta d\bar\phi,
\end{equation}
\begin{equation}
\label{eq:I5}
I_3 = \int\frac{P(\theta,\bar\phi)}{D}W^3\sin^2\theta\sin(\bar\phi) d\theta d\bar\phi.
\end{equation}
Eq.~(\ref{eq:drhodt=0}) gives the radius of the orbit,  and Eq.~(\ref{eq:omega0}) gives the angular speed of the centrosome motion with the occupancy probability $P$ determined from Eq.~(\ref{eq:dPdphi}). 
Altogether the system consists of two integrals and one differential equation for the occupancy probability $P$.

This system of nonlinear integral-differential equations are solved as follows. For an initial guess of $(\varrho^i, \omega^i)$ we solve Eq.~(\ref{eq:dPdphi}) to find the occupancy probability $P$. We compute the integrals $I's$ to update the value $(\varrho^{i+1},  \omega^{i+1})$, and compute the loss function $E = \log(1+(\delta \varrho)^2+(\delta \omega)^2)$, with $\delta \varrho = \varrho^{i+1}-\varrho^i$, and $\delta\omega=\omega^{i+1}-\omega^{i}$. We update $(\varrho, \omega)$ using an iterative gradient method for nonlinear systems to minimize the loss function within a set tolerance. 

\subsection{Numerics for the linear stability of a centrosome in a spheroid \label{subsec:Methods_arbitrary_geometry}}
Here we present numerical details for the linear stability analysis of a centrosome in a general spheroid. To obtain high-order accuracy of the numerical linear stability analysis on this differential-integral system, we need to compute surface integral
\begin{align*}
\int_{\Gamma} f({\bf Y}) dA, 
\end{align*}
with high-order spatial accuracy. We generate quadrature for spheroidal surfaces via a differential function $\varphi$, which maps $\Gamma$ to the surface of the unit sphere ($S^2$). We then transform a surface integral on $\Gamma$ to one on the unit sphere as
\begin{align}
\int_{\Gamma} \mathcal{F}({\bf Y}) dA({\bf Y}) = \int_{S^2} \mathcal{F}(\varphi({\bf X}) )  \big \lvert  \det(J_\varphi ({\bf X}) )   J_\varphi^{-t}({\bf X}) \hat{n}({\bf X}) \big \rvert  dA({\bf X}),
\end{align}
where $J_\varphi$ is the Jacobian and $\hat{n}$ is the outward normal of the sphere. This transformation allows us to efficiently compute the force on the centrosome by spatially discretizing the surface using a quadrature rule for the sphere. In particular, we put three non-overlapping patches, two caps and a ring, described by
\begin{align}
\mathcal{A}_1 &= \left \{ {\bf X} \in S^2 \, | \, \cos^{-1}({\bf X} \cdot \hat{z}) \leq \frac{\pi}{4}   \right\}, \\
\mathcal{A}_2  &= \left\{ {\bf X} \in S^2   \, | \, \frac{\pi}{4}\leq \cos^{-1}({\bf X} \cdot \hat{z}) \leq \frac{3\pi}{4}    \right\}, \\
\mathcal{A}_3 &= \left\{ {\bf X} \in S^2 \, | \,  \cos^{-1}({\bf X} \cdot \hat{z}) \geq \frac{3 \pi}{4}    \right\}.
\end{align}
Patches $\mathcal{A}_1$ and $\mathcal{A}_3$ can be discretize as a Disk, noting that $\frac{\sin(\varphi)}{\varphi}$ is smooth, hence, for the integral $\int_0^{\varphi_0} f(\varphi) \frac{\sin{\varphi}}{\varphi} \varphi  d\varphi$ we can use the same quadrature rules used for the Disk \cite{SpectralDisk}, where $\varphi$ takes the role of the radius.

\begin{figure}[t]
\includegraphics[keepaspectratio=true,scale=0.5]{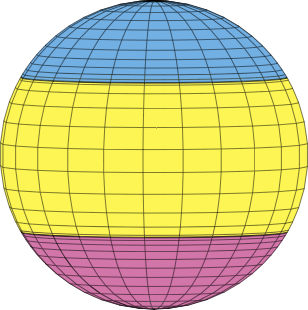}
\caption{An illustration of patched quadrature. Yellow is for the ring patch ${\cal A}_{2}$, and blue (pink) is for the top (bottom) cap patch ${\cal A}_1$ (${\cal A}_3$). Note that there is an accumulation of quadrature points around the union of the patches.
}
\label{FigPatchedQ}
\end{figure}
For $\mathcal{A}_2$ in the periodic direction a trapezoidal rule is used, on the non-periodic one a Legendre quadrature rule is used. For $\mathcal{A}_{1,3}$ we use $10 \times 10$ quadrature points, for $\mathcal{A}_2$ we use $15 \times 10$ quadrature points, periodic and height direction respectively. Fig.~(\ref{FigPatchedQ}) is an example of the three non-overlapping patches of quadrature.

\subsection{Geometric intuition of two solutions\label{subsec:Methods_geometric_intuition}}
Here we provide an intuitive argument for no more than two eigenvalues in Eq.~\ref{tras_eq}. We start with a similar equation
\begin{align}
\alpha x = - \beta + \int_{-1}^{1} \frac{\gamma }{x + t} dt = -\beta + \gamma \log \left(\frac{x+1}{x-1} \right)
\end{align}
This problem can be reduced to ask the number of solution of equation
\begin{align}
a x + b = \log \left( \frac{x+1}{x-1} \right).
\label{toy_eq}
\end{align}

\begin{figure}[t]
\includegraphics[keepaspectratio=true]{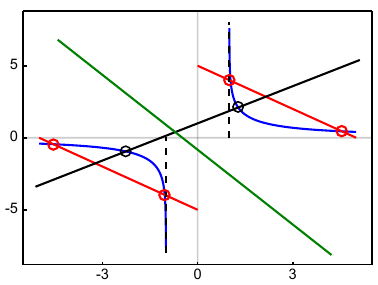}
\caption{The blue lines are the right-hand side of Eq.~(\ref{toy_eq}), and the dash lines are their two vertical asymptotes. 
The red, black and green lines are 
the left-hand side of Eq.~(\ref{toy_eq}), giving rise to three types of roots that Eq.~(\ref{toy_eq}) can have: Two roots at the intersections between the red line and the blue curve (in either the first or the third quadrant), two roots at the intersections between the black line and the blue curves, or no roots when the green curve crosses in between the asymptotes.}
\label{Fig1Sup}
\end{figure}

From simple analysis, we can see that $\log \left( \frac{x+1}{x-1} \right)$ has two vertical and one horizontal asymptotes. As we can see in Figure (\ref{Fig1Sup}) there are only two cases when there are not two real solutions. Two of them are the degenerate case, where we have that the curve is tangent to the equation of the line and the other is when the line goes through exactly between the vertical asymptotes.

\subsection{Stable singular spectrum\label{subsec:stable_singular_spectrum}}

In the main text we left out some interesting characteristics of the stable spectrum. It turns out that it has a diverse complicated structure for the slightest perturbation of the sphere, i.e. ratio $\alpha =1 \pm \epsilon$ with $0<\epsilon\ll 1$.  We start this discussion for the case when there is no centrosome dynamics, corresponding to $\bf v = 0$ in Eqs.~(\ref{eq:spheroid},\ref{eq:spheroid_p}) and the eigenvalue equation is reduced to
\begin{align}
\int_{\Gamma} p({\bf Y}) \hat{{\bf Y}} dA_{\bf Y} &= 0 \\
(\sigma + \tilde{\Omega}_0({\bf Y}) ) p({\bf Y}) &= 0 .
\end{align}
Note that if $p$ is a continuous function, the only solution to the system is $p \equiv 0$. To find the rest of the spectrum we need to consider tempered distributions. To understand this eigenfunction, we see the following simpler eigenfunction system, given by 
\begin{align} 
\lambda f(x) = (x + \epsilon)f(x) \quad x \in [-1, 1]
\end{align}
The solutions to this problem is given by the pair $(\delta(x - x_0), x_0)$, where $x_0 \in [-1, 1]$ ~\cite{MultOperator}. We can generalize this solution by thinking in cylindrical coordinates. To start, remember that all the level sets $\mathcal{A}_c = \{ {\bf Y} \in \Gamma |\;\tilde{\Omega}_0({\bf Y}) = -c \}$ from axial symmetry are given by symmetric off center circles, which leads us to define $z_c \geq 0$, such that $\tilde{\Omega}_0((0,0,z_c)) = -c$. Using these, we can conclude that if there is a singular eigenfunction of the system, its eigenvalues need to be $\sigma \in [-\max_{{\bf Y} \in \Gamma} \tilde{\Omega}_0({\bf Y}) ) ,  -\min_{{\bf Y} \in \Gamma} \tilde{\Omega}_0({\bf Y}) )]$ and its eigenfunction, $p_\sigma$, has $\text{sing supp}(p_\sigma) \subset \mathcal{A}_\sigma$. With this idea in mind we can determine the eigenvalue spectrum $\{\sigma\} = [-\max_{{\bf Y} \in \Gamma} \tilde{\Omega}_0({\bf Y}) )  ,  -\min_{{\bf Y} \in \Gamma} \tilde{\Omega}_0({\bf Y}) )]$, with the corresponding generalized eigenfunctions
\begin{align}
p_{\sigma}^{n, \pm}(z, \theta) =
\begin{cases}
e^{i n \theta} \delta(z \pm z_c)\quad n \in \mathbb{Z},\, |n| \geq 2 \\
e^{\pm i \theta} ( \delta(z - z_c) - \delta(z + z_c)   )  \quad n = 1  \\
\delta(z - z_c) + \delta(z + z_c) \quad n = 0
\end{cases}
\label{Eq:EigenDelta}
\end{align}

\begin{figure}[h]
\includegraphics[scale = 0.25]{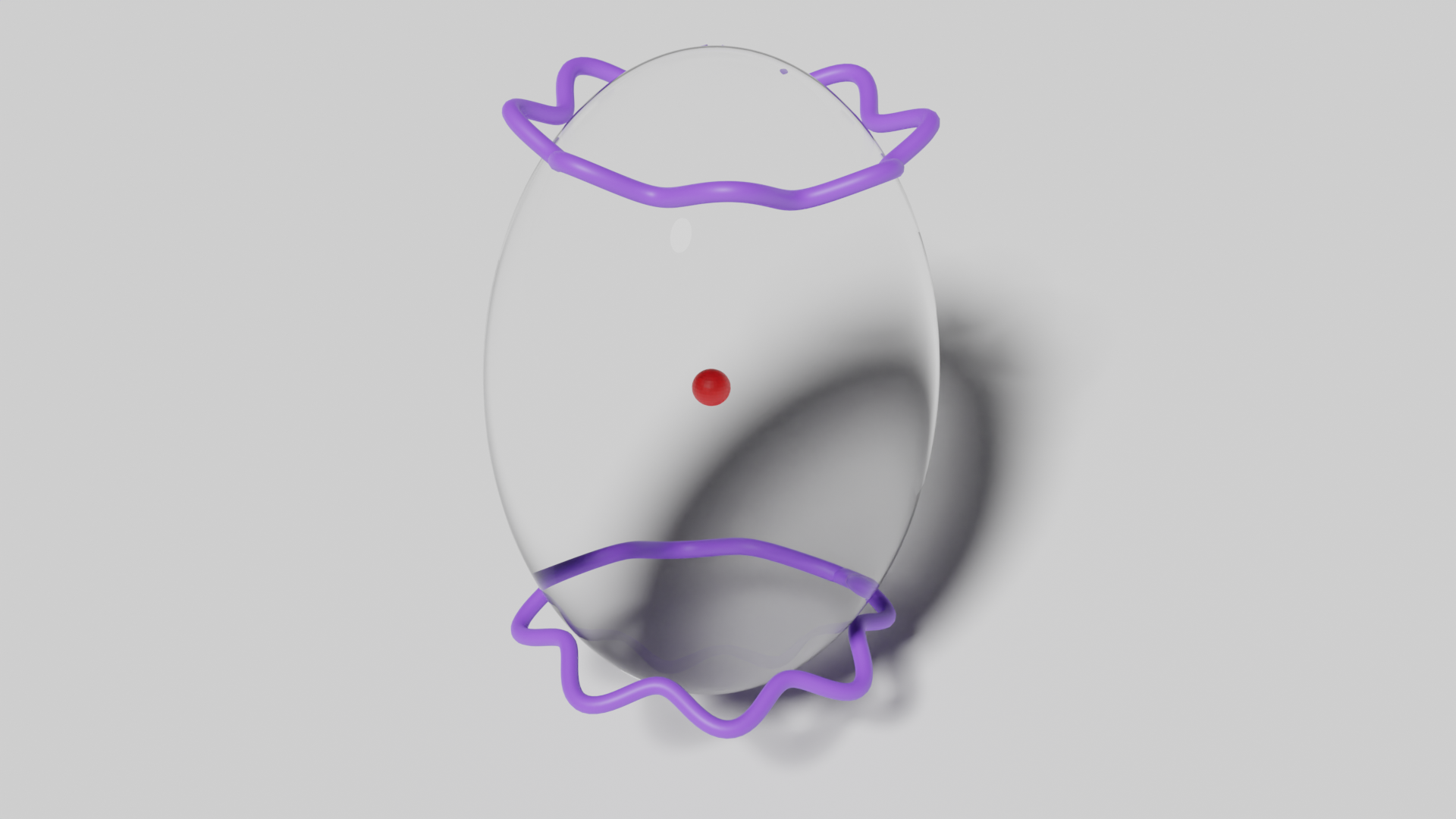}
\caption{Graphical representation of $p_{\sigma}^{n, \pm}(z, \theta)$ for a spheroid with $\alpha = 1.5$. The curves over the surface represent the eigenfunction that is either 0 or infinity times a trigonometric function, as in Eq.~(\ref{Eq:EigenDelta}). The red sphere denotes the stably centered aster.
}
\label{fig:EasterEgg}
\end{figure}
A schematic of the generalized eigenfunctions is shown in Fig.~(\ref{fig:EasterEgg}). We remark how much the structure of the eigenvectors change with a perturbation of the sphere. For the sphere, there was only one eigenvalue associated with modes that do not generate centrosome motion. Its corresponding eigenspace could be decomposed into a countable set of spherical harmonics. Meanwhile, for the ellipsoidal case, we have a continuum spectrum of modes that do not generate centrosome motion, where each eigenspace is spanned by a countable set of generalized functions.

Incredibly, we find more structure of the solution when we depart from the standard analysis and reinterpret the integrals in Eq.~(\ref{tras_eq_simple}) as principal value integrals, opening the possibility to find new nonlinear solutions of the equation in the range $[-\max_{{\bf Y} \in \Gamma} \tilde{\Omega}_0({\bf Y}) ),  -\min_{{\bf Y} \in \Gamma} \tilde{\Omega}_0({\bf Y}) ]$. Numerically we can find there is at least one solution for each index on the parameter range that we studied. That being said, the corresponding eigenvectors are decaying modes and take singular values, hence not of physical interest.

\subsection{Probabilistic derivation of the overlap model\label{subsec:overlap_model}}
Here we provide detailed derivations of the exponential model for the interaction between force generators and the astral MTs. We present
two different approaches. One is a direct, explicit derivation with all the hypothesis of the problem and substantial algebraic details,  and the other simpler calculation making an analogy to picking balls of different colors,  with less details. To simplify the notation, we introduce the following events
\begin{itemize}
\item ${\bf A}_{\bf Y} = $ MT hits CFG$_{\bf Y}$ (CFG centered at point ${\bf Y}$)
\item ${\bf B}_{\bf Y}=$ MT binds to CFG$_{\bf Y}$
\item ${\bf C}_{\bf Y} = $ CFG$_{\bf Y}$ is occupied
\item ${\bf D}^n_{\bf Y} = $ There are $n$ other CFG competing for the same MT that already binds to a CFG$_{\bf Y}$
\end{itemize}

Note that in this notation $P({\bf Y}) = \mathbb{P}({\bf C}_{\bf Y})$. 

Let $U$ be a patch over the surface $\Gamma$, centered at $\bf Y$, of area $A_U$ such that $\forall {\bf X} \in U$, $|P({\bf X}) - P({\bf Y})| << 1$. In the patch the total number of force generators is given by $N_U = M \int_{U} \rho({\bf X}) dA \approx M A_U\rho({\bf Y}) $. To start the derivation we see that the change in $P({\bf Y})$ in a time-step $\Delta t$ is given by
\begin{align}
d P({\bf Y}) = - \kappa P({\bf Y}) \Delta t + \mathbb{P}(\text{MT binds to CFG$_{\bf Y}$ in $\Delta t$}).
\end{align}
By the law of total probabilities, we have 
\begin{align}
\mathbb{P}(\text{MT binds to CFG$_{\bf Y}$ in $\Delta t$}) = \Omega({\bf Y}) \,  \mathbb{P}({\bf B}_{\bf Y} | {\bf A}_{\bf Y})\,\Delta t.
\end{align}
The details of modeling CFG will be expressed through the term $\mathbb{P}({\bf B}_{\bf Y} | {\bf A}_{\bf Y})$. For stochiometric interactions, i.e., a free astral MT cannot bind to a CFG that is already occupied, we have $\mathbb{P}({\bf B}_{\bf Y} | {\bf C}_{\bf Y} ) = 0 $. Therefore, by the law of total probabilities, we have
\begin{align}
\mathbb{P}({\bf B}_{\bf Y} | {\bf A}_{\bf Y} ) = (1 - P({\bf Y}) ) \mathbb{P}({\bf B}_{\bf Y} | {\bf A}_{\bf Y}, {\bf C}_{\bf Y}^c ) 
\end{align}

We first note that if there was no interaction with other CFGs, then  $\mathbb{P}({\bf B}_{\bf Y} | {\bf C}_{\bf Y}^c ) = 1$, yielding the independent motor model. However in this work we consider situations where there can be overlap between CFGs.  To be specific we consider that if a MT hits a point where there is an overlap between $n$ unoccupied CFGs, this MT will bind randomly to only one of these. In terms of probabilities, it means that $\mathbb{P}({\bf B}_{\bf Y} | {\bf A}_{\bf Y},{\bf C}_{\bf Y}, {\bf D}_{\bf Y}^n  ) = \frac{1}{1 + n}$. Now, for a CFG$_{\bf X}$ to be available to compete for the MT it needs to be unbound and there has to be overlap between CFG$_{\bf X}$ and CFG$_{\bf Y}$. Since these two events are independent, we just need to approximate the probability that there is overlap between different CFGs in the patch $U$. For this we do a simple homogeneity approximation. From here we can see that $D_{\bf Y}^n$ is a binomial distribution (in $n$) of parameters $(N_U-1, \frac{a}{A_u}(1 - P(Y)))$, where $a=\pi r_m^2$ is the discal area of the CFG binding domain. In other words we have
$$\mathbb{P}({\bf D}^n_{\bf Y}| {\bf A}_{\bf Y}) = {N_U -1 \choose n} \left((1-P) \frac{a}{A_u} \right)^n \left(1 - (1-P) \frac{a}{A_u} \right)^{N_U -1 - n} $$
Putting these together using the law of total probabilities, we then have to evaluate the sum 
\begin{align}
\mathbb{P}({\bf B}_{\bf Y}| {\bf A}_{\bf Y}) = (1 - P({\bf Y})) \sum_{n = 0}^{N_U -1}\frac{1}{1+n} {N_U -1 \choose n} \left((1-P) \frac{a}{A_u} \right)^n \left(1 - (1-P) \frac{a}{A_u} \right)^{N_U -1 - n}.
\end{align}
This sum can be easily evaluated by first letting  $q = (1 - P(\mathbf{Y})\frac{a}{A_U})$ and $J = N_U-1$, then we have
\begin{align}
\sum_{n=0}^J {J \choose n} \frac{q^n}{n+1}(1-q)^{J-n} =\frac{1}{q} \int_0^q  \sum_{n=0}^J {J \choose n}  x^n (1-q)^{J-n}dx . 
\end{align}
Using Newton's binomial theorem and integrating, we obtain the final expression
\begin{align}
\mathbb{P}({ \bf B}_{\bf Y}| {\bf A}_{\bf Y}) = \frac{1}{N_U  \frac{a}{A_u}} \left(  1 - (1 - (1-P) \frac{a}{A_u})^{N_U}.  \right)
\end{align}
From this expression, we obtain the following three interesting results.
\begin{enumerate}
\item If there is no overlap between CFGs, $A_u = a$ and $N_U = 1$, and we recover the independent CFG model where 
\begin{align}
\mathbb{P}({\bf B}_{\bf Y}| {\bf A}_{\bf Y}) = 1- P
\end{align}
\item Taking the limit when $\frac{a}{A_u} \to 0$ we obtain the interaction probability ${\cal I}(P)$ used in this work
\begin{align}
\mathbb{P}({\bf B}_{\bf Y}| {\bf A}_{\bf Y}) = \frac{1 - e^{- aM \rho(1-P)}}{aM \rho} = {\cal I}(P).
\label{GenerealProb}
\end{align}
\item If the CFGs are perfectly stacked one over each other (instead of being randomly distributed), this means that $N_U$ is independent of $a$, so we can just call it $N$ and we get
\begin{align}
\mathbb{P}({\bf B}_{\bf Y}| {\bf A}_{\bf Y}) = \frac{1 - P^N}{N}
\end{align}
\end{enumerate}

Below we present a second derivation, simpler in the algebra, that hides many of the details of the interactions. First we define the event $${\bf E}_U = \text{MT hits patch } U,$$  using law of total probabilities we have  
\begin{align}
\mathbb{P}({\bf B}_{\bf Y} |{\bf E}_U) = \mathbb{P}({\bf B}_{\bf Y} |{\bf E}_U, {\bf A}_{\bf Y}) \mathbb{P}({\bf A}_{\bf Y}|{\bf E}_U) + \mathbb{P}({\bf B}_{\bf Y} |{\bf E}_U, {\bf A}_{\bf Y}^C) \mathbb{P}({\bf A}_{\bf Y}^C|{\bf E}_U).
\end{align}

The probability of binding to CFG$_{\bf Y}$ with no MT hitting the area that it covers, is 0. Further, conditioning over ${\bf E}_U$ and ${\bf A}_{\bf Y}$ is redundant, since the latter implies the former. Thus the expression is simplified to 
\begin{align}
\mathbb{P}({\bf B}_{\bf Y} |{\bf E}_U) = \mathbb{P}({\bf B}_{\bf Y} |{\bf A}_{\bf Y}) \mathbb{P}({\bf A}_{\bf Y}|{\bf E}_U).
\end{align}

We are interested in (re)determining $\mathbb{P}({\bf B}_{\bf Y} |{\bf A}_{\bf Y})$, but $\mathbb{P}({\bf B}_{\bf Y} |{\bf E}_U)$ has the advantage that every CFG within patch $U$ can be treated equally. To simplify the computation of this quantity, we set up an equivalent but simpler problem. Take $N_U$ balls, with one of them being red, and put them in two different boxes, each ball with probability $(1-P)\frac{a}{A_u}$ being in box one,  and in the other box with the complement probability. Next, one ball is picked from box one (if there is any ball). We pose the question: What is the probability that the red ball is picked? 

It is not difficult to see that getting the red ball has a probability of $\mathbb{P}({\bf B}_{\bf Y} |{\bf E}_U)$. In this setting, we just need to determine the probability that there is at least one ball in box one and that the picked ball is red. Since all of the balls are the same, these two probabilities are independent, so it is straightforward that
$$\mathbb{P}({\bf B}_{\bf Y} |{\bf E}_U) = \frac{1}{N_U} \left(  1 - (1 - (1-P) \frac{a}{A_u})^{N_U}  \right).$$
Noting that $\mathbb{P}({\bf A}_{\bf Y}|{\bf E}_U) = \frac{a}{A_U}$ finalizes the proof.

\subsection{Aster dynamics inside a cuboid\label{sec:ForceMap_Microchamber}}
\begin{figure}[t]
\includegraphics[keepaspectratio=true]{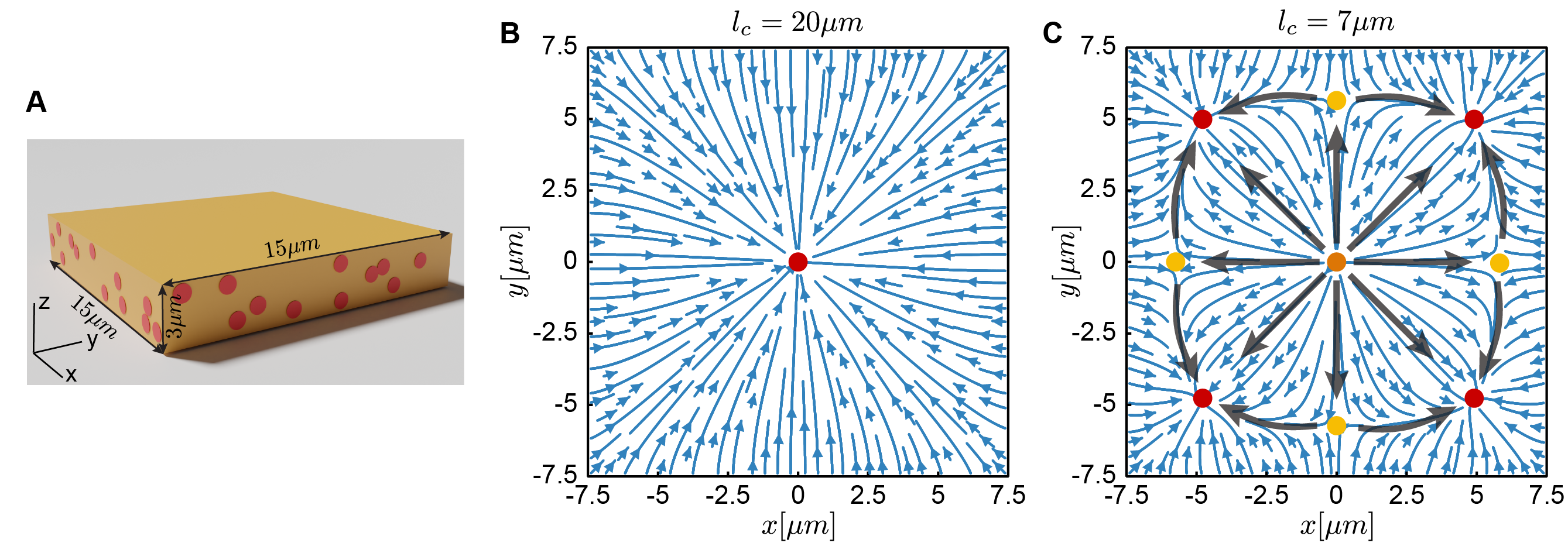}
\caption{Distribution of pulling force on a centrosome in a square slab-like cell. (A) A square chamber with CFGs on the sidewall, similar to the microfabricated chamber with dynein motors attached on the sidewalls in \cite{laan2012cortical}. (B)-(C) By computing the force required to fix the centrosome at a point inside the chamber, 
%{\bf MJS: How are these calculated?} 
we obtain the direction fields of the pulling force on the centrosome inside a square chamber for characteristic MT length of $l_c=20 \mu m$ and $7 \mu m$, respectively. Red circles are for stable fixed points (where the net forces converge), yellow circles are for saddle fixed points (where the net forces converge in one direction and diverge in another), and orange circles denote unstable fixed points (where the net forces diverge). 
%{\bf MJS: Is stability the property that should be connoted here?}
%(E)-(F) Oscillation period measured at $M=M_H$ for the supercritical Hopf bifurcation for parameter values in Table~\ref{tab:parameters} for three values of $V_g=1, 0.5, 1.5 \mu m/s$.
}
\label{Fig9}
\end{figure}

We conducted simulations of the S-model with a centrosome placed inside a cuboid of dimensions $15\times 15\times 3\;\mu m$ with CFGs positioned exclusively on the peripheral sides (see Fig.~\ref{Fig9}A), imitating the microchamber in the experiments \cite{laan2012cortical, ma2014general, pavin2012positioning}. For a given characteristic MT length, we compute the pulling force distribution minus the force needed to keep the centrosome at a given point inside the chamber. When the average MT length ($l_c=20\;\mu m$) is comparable to the length of the microchamber, a singular fixed point emerges at the chamber's center (Fig.~\ref{Fig9}B). For shorter MTs, a constellation of fixed points appears, distributed around the central region (Fig.~\ref{Fig9}C). The central fixed point becomes unstable, while eight additional fixed points formed, with those near the chamber's corners exhibiting stability (red in Fig.~\ref{Fig9}C), and the intermediates acting as saddle points (yellow in Fig.~\ref{Fig9}C). Previous theoretical models successfully explained the positioning of the aster in these microchambers by a balance of pulling forces from motors and pushing forces from MT growth and friction with the chamber periphery \cite{laan2012cortical, ma2014general, pavin2012positioning}. Our findings provide an alternative explanation for the positioning of centrosomes in microchambers, suggesting it results from the combined effect of pulling forces and the stoichiometric interactions between MTs and molecular motors.

\end{document}